\documentclass[final]{agujournal2019}
\makeatletter
\def\@oddhead{}
\def\@evenhead{}
\makeatother
\graphicspath{{figures/}} % committed figures live in paper/figures/ (FIGURES.md); LaTeX still falls back to the current directory
\usepackage{url} %this package should fix any errors with URLs in refs.
\usepackage{amsmath}
\usepackage{amssymb}
\usepackage{siunitx}

\journalname{}

\begin{document}

%%%%%%%%%%%%%%%%%%%%%%%%%%%%%%%%%%%%%%%%%%%%%%%
%  TITLE
%
% (A title should be specific, informative, and brief. Use
% abbreviations only if they are defined in the abstract. Titles that
% start with general keywords then specific terms are optimized in
% searches)
%
%%%%%%%%%%%%%%%%%%%%%%%%%%%%%%%%%%%%%%%%%%%%%%%

% Example: \title{This is a test title}

\title{Extremes on Rewind: Generating 1,000-Member Ensembles Initialized at a Final Condition}

%%%%%%%%%%%%%%%%%%%%%%%%%%%%%%%%%%%%%%%%%%%%%%%
%
%  AUTHORS AND AFFILIATIONS
%
%%%%%%%%%%%%%%%%%%%%%%%%%%%%%%%%%%%%%%%%%%%%%%%

% Authors are individuals who have significantly contributed to the
% research and preparation of the article. Group authors are allowed, if
% each author in the group is separately identified in an appendix.)

% List authors by first name or initial followed by last name and
% separated by commas. Use \affil{} to number affiliations, and
% \thanks{} for author notes.
% Additional author notes should be indicated with \thanks{} (for
% example, for current addresses).

\authors{Jerry Lin$^{1}$, Mu-Ting Chien$^{1}$, Mansi Sakarvadia$^{2}$, Elizabeth A. Barnes$^{1}$}

% % \affiliation{1}{First Affiliation}
% % \affiliation{2}{Second Affiliation}
% % \affiliation{3}{Third Affiliation}
% % \affiliation{4}{Fourth Affiliation}

\affiliation{1}{Department of Computing \& Data Sciences, Boston University, Boston, MA, USA}
\affiliation{2}{Department of Computer Science, University of Chicago, Chicago, IL, USA}
% %(repeat as many times as is necessary)

% % Corresponding author mailing address and e-mail address:

% % (include name and email addresses of the corresponding author.  More
% % than one corresponding author is allowed in this LaTeX file and for
% % publication; but only one corresponding author is allowed in our
% % editorial system.)

% % Example: \correspondingauthor{First and Last Name}{email@address.edu}

% \correspondingauthor{Jerry Lin}{jlin404@bu.edu}

%%%%%%%%%%%%%%%%%%%%%%%%%%%%%%%%%%%%%%%%%%%%%%%
% KEY POINTS
%%%%%%%%%%%%%%%%%%%%%%%%%%%%%%%%%%%%%%%%%%%%%%%
%  List up to three key points (at least one is required)
%  Key Points summarize the main points and conclusions of the article
%  Each must be 140 characters or fewer with no special characters or punctuation and must be complete sentences

% Example:
% \begin{keypoints}
% \item	List up to three key points (at least one is required)
% \item	Key Points summarize the main points and conclusions of the article
% \item	Each must be 140 characters or fewer with no special characters or punctuation and must be complete sentences
% \end{keypoints}

\begin{keypoints}
\item We use cBottle-video to directly sample candidate histories of chosen extreme events as 1000-member ensembles of 66-hour weather sequences.
\item Generated histories of a given extreme are nearly as diverse as forward forecasts, producing 84 to 89 percent of their 500 hPa height spread.
\item Antecedent height patterns explain 48 percent of tracked Sandy latitude variance but under 10 percent of PNW heatwave temperature spread.
\end{keypoints}

\begin{abstract}

Scenario planning for rare, high-impact events often requires massive ensembles to stochastically sample relevant trajectories. Although autoregressive weather emulators can efficiently generate such ensembles, isolating trajectories of interest requires sifting through petabytes of data, a challenge that grows with lead time and rarity. In contrast, a non-autoregressive foundation model like Climate in a Bottle video (cBottle-video) can directly sample trajectories terminating in extremes, avoiding large-ensemble search. We use cBottle-video to generate 1000-member ensembles with start- and/or final-conditioning across three extreme events: the 2021 Pacific Northwest (PNW) heatwave, Superstorm Sandy, and Hurricane Ian. The antecedent 500 hPa geopotential height ($z_{500}$) spread in final-conditioned ensembles is 84--89\% as large as the forecast spread in start-conditioned runs, suggesting diverse pathways to each extreme. For the 2021 PNW heatwave, final-conditioned ensemble members begin uniformly warmer than reanalysis and stay warm, replacing the observed rapid intensification with persistent antecedent heat. For Superstorm Sandy, the leading modes of $z_{500}$ at the antecedent end of the final-conditioned ensemble explain 48\% of the variance in track latitude for ensemble members whose storm is detected at the first frame, and about 11\% of those members begin with lower central pressure than Sandy. For Hurricane Ian, variation in the first landfall location among final-conditioned trajectories underscores the potential for exploring diverse intermediate hazard exposure.

\end{abstract}

\section*{Plain Language Summary}

To prepare for extreme weather events, planners need to know how such events might unfold. Current methods for anticipating this involve running weather models many times from slightly different starting conditions and checking whether any of the resulting simulations produce the extreme event of interest---an inefficient process that becomes harder the longer the simulation and the rarer the event. In this paper, we introduce a fundamentally different approach, using an artificial intelligence model from the ``Climate in a Bottle'' family that generates entire multi-day weather sequences at once, rather than stepping forward in time. With this model, we can generate a thousand 66-hour weather sequences that all end in the same extreme event. We apply this method to three test cases: the 2021 Pacific Northwest heatwave, Superstorm Sandy, and Hurricane Ian. The model generates very different paths that all end in the same extreme event, although not all of them are necessarily physically realistic. If approaches like this eventually prove to be robust, they could give emergency managers and policymakers a powerful new tool: the ability to ask ``what kinds of weather patterns could lead to \textit{this} disaster?'' and get a thousand plausible answers.

%%%%%%%%%%%%%%%%%%%%%%%%%%%%%%%%%%%%%%%%%%%%%%%
%
%  BODY TEXT
%
%%%%%%%%%%%%%%%%%%%%%%%%%%%%%%%%%%%%%%%%%%%%%%%
\section{Introduction}

Data-driven autoregressive weather emulators and end-to-end differentiable hybrid models with machine learning components have revolutionized the field of weather forecasting, generating forecasts faster and at lower computational cost than traditional physics-based models \cite{Bi2023-bg, Lam2023-jq, Kochkov2024-uy, Watt-Meyer2025-oj, Bonev2025-dw, Davenport2026-ro, Yuval2026-pk, Gentine2026-ml}. While purely data-driven models cannot replace physics-based models given deficiencies in mechanistic interpretability and generalizability out-of-distribution \cite{Sun2025-mh, Craig2026-ex, Hover2026-wv, Merchant2026-ex}, their ability to quickly generate huge ensembles makes them especially useful for sampling low-likelihood, high-impact events \cite{Mahesh2025-py, Mahesh2025-ue}. Unfortunately, generating and storing large ensembles can produce substantial data volumes. As a concrete example, the huge ensemble (HENS) dataset created by \citeA{Mahesh2025-ue} comprises 7,424-member ensembles initialized on each of 92 days. With 15-day rollouts saved at $0.25^{\circ}$ resolution across 12 variables, the data occupies approximately two petabytes of storage. This problem only becomes more severe with greater lead times and increasingly extreme events.

Nevertheless, the flexible design and end-to-end differentiability of this new generation of data-driven and hybrid weather and climate models expand the range of possible experiments. For example, end-to-end differentiability allows for the direct optimization of initial conditions to guide the state of the atmosphere towards a desired condition \cite{Whittaker2026-fe, Hakim2026-oy}. This has been used to amplify the intensity of the 2021 Pacific Northwest (PNW) heatwave in \citeA{Whittaker2026-fe} and alter the trajectory of Hurricane Fiona to more closely align with the anomalous track of Superstorm Sandy in \citeA{Hakim2026-oy}. That said, the optimization process perturbs these initial conditions only slightly by design, exploring a narrow neighborhood of the original state rather than the full distribution of initial conditions consistent with a given extreme event. Fortunately, the arrival of climate foundation models like Climate in a Bottle video (cBottle-video) has introduced a fundamentally different paradigm by allowing for the generation of weather sequences that share a final, rather than initial, condition \cite{Brenowitz2025-mz}. This is because cBottle-video, a video variant of the original cBottle de-noising diffusion model, generates twelve frames corresponding to a 66-hour weather sequence in parallel, with the option to condition an arbitrary subset of these frames on existing data \cite{Brenowitz2025-mz}. This inverts the usual forecasting question: rather than asking what an initial state produces when time is run forward, one can instead ask which antecedent conditions could culminate in a given extreme by directly generating candidate histories for the event, circumventing the need to filter them from a much larger ensemble.

In this work, we present the first application of final-conditioned video diffusion to directly sampling candidate histories of chosen extreme events. Using the flexible conditioning capability of cBottle-video, we generate 1000-member ensembles across three historical case studies: the 2021 PNW heatwave, Superstorm Sandy, and Hurricane Ian. For completeness, we also generate 1000-member ensembles conditioned on the start frame (analogous to conventional weather forecasts) and conditioned at both the final and start frame (resulting in a form of interpolation between two atmospheric states). Given that we are evaluating an off-the-shelf checkpoint that was not fine-tuned for creating diverse yet physically realistic final-conditioned ensembles, we do not expect the final-conditioned trajectories to be free of physical inconsistencies or to be inherently dynamically admissible. Instead, we seek to evaluate whether final-conditioning can sample diverse antecedent pathways for extreme events without large-ensemble search and to diagnose areas where concrete physical inconsistencies arise to motivate directions for future model development.

\section{Model Description and Methods}

cBottle-video uses an adapted Song-UNet diffusion backbone augmented with a learned position embedding, time of day and time of year embedding, monthly mean sea surface temperature (SST) conditioning, and temporal self-attention layers \cite{Song2020-hs, Karras2022-xq, Ho2022-lf, Blattmann2023-zo, Brenowitz2025-mz}. To avoid geometric issues associated with getting convolutional filters to learn location-invariant patterns on a sphere, the model operates on a HEALPix (HPX64) equal-area spherical discretization with custom padding to allow for the flow of information between adjacent faces \cite{Karlbauer2024-nv}. In the HEALPix grid, which corresponds to 100 km resolution, the surface of the Earth is divided into 12 square faces of $64\times64=4{,}096$ pixels each (49,152 pixels globally), and the 64 in HPX64 denotes the number of pixels along one edge of one face \cite{Karlbauer2024-nv, Brenowitz2025-mz}.

The model is trained using a score-based diffusion framework and a masked-conditioning training scheme in which mask patterns (i.e. random frame dropout, block-wise masking, endpoint interpolation, and full sequence masking) are randomly sampled during training to allow the model to infill arbitrary temporal gaps within a 66-hour (12 frame) weather sequence at 6-hour resolution \cite{Song2020-hs, voleti2022MCVD, Brenowitz2025-mz}. The model's training objective is shown in Equation \ref{eq:training_objective}. Samples are generated by solving the reverse-time differential equation shown in Equation \ref{eq:sampling_step} using the deterministic second-order Heun sampler of \citeA{Karras2022-xq} ($S_{\text{churn}} = 0$). 

\begin{equation}
\label{eq:training_objective}
\arg\min_{\theta} \mathbb{E}_{\mathbf{x},\mathbf{c}} \mathbb{E}_{\sigma \sim p_\sigma} \mathbb{E}_{\epsilon \sim \mathcal{N}(0, \sigma^2 \mathbf{I})} \left[ \lambda(\sigma) \| \mathcal{D}_\theta(\mathbf{x} + \epsilon; \sigma, \mathbf{c}) - \mathbf{x} \|^2 \right], \quad \lambda(\sigma) = \frac{\sigma^2 + \sigma_{\text{data}}^2}{\sigma^2\sigma_{\text{data}}^2}
\end{equation}

\begin{equation}
\label{eq:sampling_step}
\frac{d\mathbf{x}_\sigma}{d\sigma} = \frac{\mathbf{x}_\sigma - \mathcal{D}_\theta(\mathbf{x}_\sigma; \sigma, \mathbf{c})}{\sigma}
\end{equation}

Here, $\mathbf{x}$ is a 12-frame sequence of the model's 45 atmospheric and surface variables on the HPX64 grid, each standardized by its global mean and standard deviation so that $\sigma_{\text{data}} = 1$, and $\mathbf{c}$ is the conditioning (i.e. unmasked frames, frame mask, SST, time of day and year, and dataset label selecting ERA5 rather than ICON simulation output as the target distribution). During training, Gaussian noise $\epsilon$ with standard deviation $\sigma$ (the noise level) is added to $\mathbf{x}$, and the denoiser $\mathcal{D}_\theta$, a U-Net with weights $\theta$, is trained to recover $\mathbf{x}$ from the noisy sequence given $\sigma$ and $\mathbf{c}$ by minimizing the expected squared error. The weight $\lambda(\sigma)$ balances the contributions of different noise levels to this loss \cite{Karras2022-xq}. The training noise distribution $p_\sigma \propto \sigma^{-1}$ is log-uniform between $\sigma_{\min} = 0.02$ and $\sigma_{\max} = 1000.0$, following the spectral criterion of \citeA[Section A.3.1]{Brenowitz2025-mz}. According to this criterion, $\sigma_{\max}^2$ should exceed the variance of the leading principal component of the data so that the noise swamps even the largest-scale, highest-amplitude signals such as the diurnal and seasonal cycles. Because the variance of this leading mode grows with the number of frames over which such signals remain correlated, the 12-frame video model requires a larger $\sigma_{\max}$ than the value of 200 used for the original single-frame cBottle model \cite[Section A.8.2]{Brenowitz2025-mz}. In Equation \ref{eq:sampling_step}, $\mathbf{x}_\sigma$ is the noisy sequence being generated at noise level $\sigma$, playing the role of $\mathbf{x} + \epsilon$ in Equation \ref{eq:training_objective}, and $\mathcal{D}_\theta(\mathbf{x}_\sigma; \sigma, \mathbf{c})$ is the denoiser's estimate of the corresponding clean sequence, so the numerator is the estimated noise remaining in $\mathbf{x}_\sigma$. Integration starts from pure Gaussian noise with standard deviation $\sigma_{\max}$ and removes this estimated noise as $\sigma$ decreases to zero, where $\mathbf{x}_\sigma$ becomes a draw from the learned distribution of sequences given $\mathbf{c}$. Because this equation is deterministic, ensemble members differ only in their initial noise (Text S1). Further details regarding the model's architecture and training procedure can be found in \citeA[Sections \mbox{A.7--A.8}]{Brenowitz2025-mz}.

For our experiments, we load a pre-trained checkpoint of cBottle-video that is publicly available on Hugging Face using a modified version of earth2studio, an open-source software repository designed to streamline the inference and evaluation of deep-learning weather emulators \cite{Earth2Studio_Contributors_NVIDIA_Earth2Studio_2024}. Because implementations of different emulators are standardized under a common API, the original implementation of cBottle-video functions as an autoregressive model by default, and our fork of earth2studio subclasses this implementation to unlock the model's existing flexible conditioning capability. All ensembles generate 12 frames at a time, but final-conditioned ensembles are conditioned on the final frame, start-conditioned ensembles are conditioned on the first frame, and start-and-final-conditioned ensembles are conditioned on both the start and final frames. Text S1 details the sampler settings and the seeding of the 1000 members. All conditioning frames and ground-truth evaluations make use of the ERA5 reanalysis dataset, which is provided at $0.25^{\circ}$ resolution on a lat-lon grid and regridded to HPX64 prior to being used for conditioning \cite{Hersbach2020-oh}. SST conditioning instead uses AMIP mid-month boundary-condition values interpolated to each frame's valid time, prescribed rather than coupled and identical across members \cite{Brenowitz2025-mz}. Further details regarding AMIP SST conditioning can be found in Text S1. Tropical cyclone tracks are extracted at 6-hourly intervals using the TempestExtremes tracker \cite{Ullrich2021-nm} applied to the $0.25^{\circ}$-regridded fields. Candidate storm centers are detected as minima of the mean sea level pressure (MSLP) field subject to a closed-contour criterion and the warm-core thickness criterion of \citeA{Zarzycki2017-fo}, then stitched into tracks subject to persistence, wind-speed, and latitude thresholds. The full tracker configuration is detailed in Text S3. Because the $0.25^{\circ}$ grid adds no information beyond the model's native HPX64 ($\sim$100~km) resolution, storm-center fixes are effectively resolved at that coarser scale.

Given the fact that synoptic drivers for low-likelihood, high-impact events often extend a week or more prior, it is tempting to extend the ``rewound'' trajectories past 66 hours by running cBottle-video autoregressively (i.e. conditioning additional 66-hour trajectories using the endpoints of previously generated trajectories) \cite{White2023-fd}. However, longer trajectories into the past are likely better suited for follow-up work as the checkpoint used in the experiments shown here was not trained for, or evaluated on, autoregressive performance \cite{Brenowitz2025-mz}. Moreover, because conditioned frames enter the model only through the conditioning $\mathbf{c}$ and are regenerated rather than copied into the output, they are reproduced with subtle but noticeable degradations (see Figures~S15--S17), motivating the development of methods that can mitigate compounding degradation in autoregressive rollouts, the training of a new video diffusion model with a context window that extends past 66 hours, or an autoregressive emulator that is trained from scratch to go backwards in time \cite{Hassanzadeh2026-dd}.

Across our three case studies, we examine whether variation among generated histories is statistically organized by the dominant regional midtropospheric height patterns. We characterize these patterns through EOF analysis of the $z_{500}$ field at the unconditioned boundary, using a fixed regional subdomain for each case. This provides a consistent diagnostic across cases, motivated by our hypothesis that the severity of the heatwave as well as the storm position and intensity of the hurricanes should co-vary with regional midtropospheric geopotential height patterns. We quantify this organization using the 10-fold cross-validated $R^2$ of multiple linear regressions predicting impact-domain land-mean 2-meter temperature ($T_{\mathrm{2m}}$) for the heatwave, or storm latitude, longitude, and central mean sea level pressure (MSLP) for the hurricanes, from the leading eight regional $z_{500}$ principal component (PC) scores (Text S2).

\section{Results}

\subsection{2021 Pacific Northwest heatwave}

\begin{figure}
    \centering
    \includegraphics[width=\textwidth]{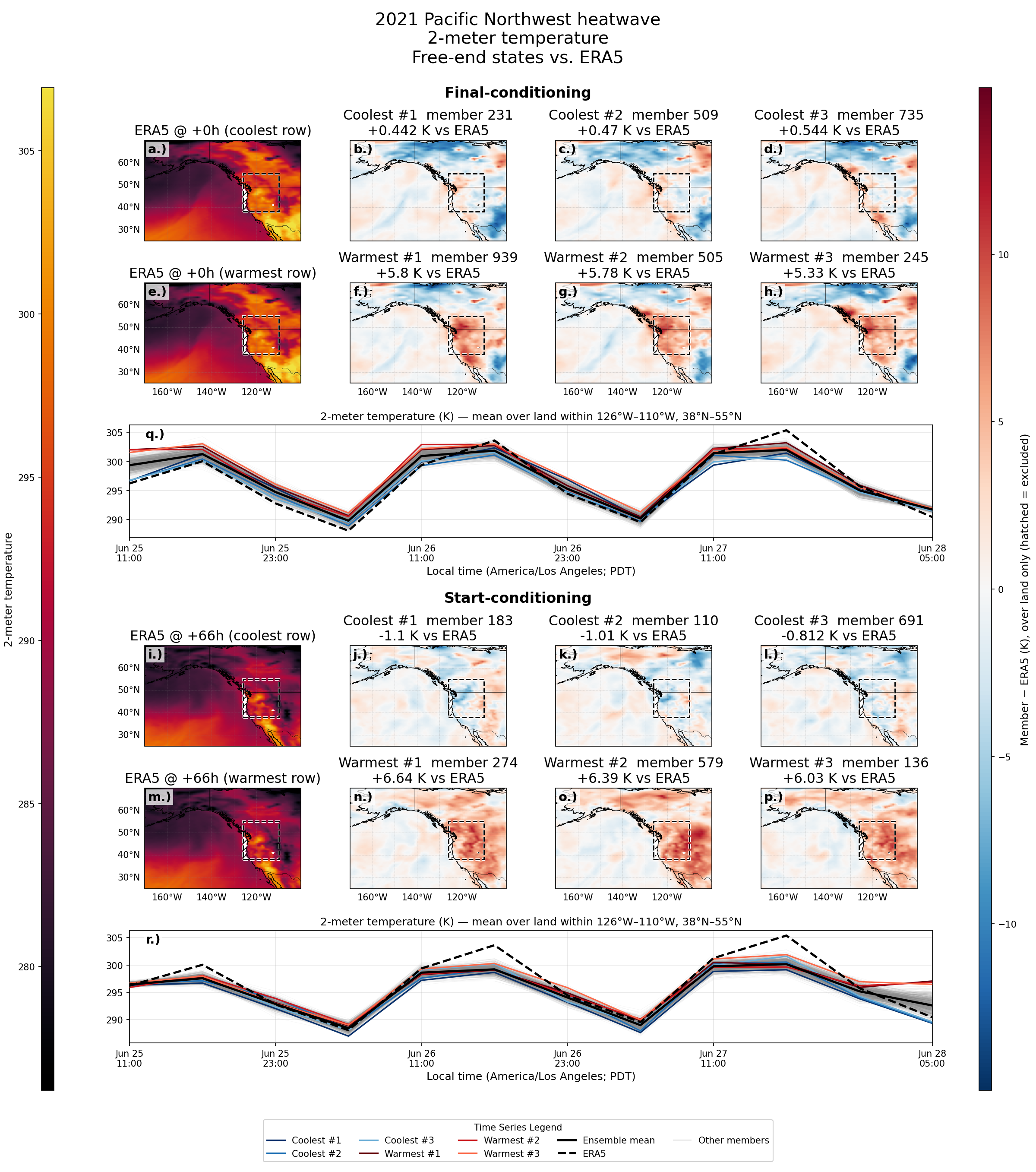}
 \setlength{\belowcaptionskip}{-1em}%

\caption{For each conditioning set-up, the top two rows show the 2-meter temperature difference (member - ERA5) for the three warmest and three coolest ensemble members, with a land mask applied over the impact domain ($126^{\circ}$–$110^{\circ}$ W, $38^{\circ}$–$55^{\circ}$ N); the leftmost panel of each row is the ERA5 reference. The third row shows the trajectories of the mean 2-meter temperature over the same domain, with the aforementioned warmest and coolest members shown in red and blue, respectively.}

 \label{fig:free_end_states_pnw_heatwave}
\end{figure}

In late June 2021, an extraordinary heatwave struck the Pacific Northwest, breaking all-time temperature records and causing hundreds to over a thousand excess deaths across the region \cite{Neal2022-wk, White2023-fd, Mass2024-zq}. The event was driven by a multi-day, strongly amplified 500~hPa geopotential height ($z_{500}$) ridge whose growth was reinforced by diabatic heating within an upstream North Pacific cyclone \cite{Neal2022-wk, Schumacher2022-gb}. Beneath the ridge, record surface temperatures arose from a combination of near-maximal solstice insolation under clear skies, adiabatic warming from enhanced subsidence and offshore flow, and reduced cooling as the Pacific marine layer was suppressed \cite{Neal2022-wk, Schumacher2022-gb, White2023-fd, Loikith2023-nt, Mass2024-zq, Ndoping2026-sc}. Although past heat waves shared similar ingredients, the June 2021 event was distinguished by the exceptional magnitude of several key dynamical factors, including record-breaking midtropospheric heights and temperatures and stronger thermal troughing/offshore flow than historical analogues \cite{Loikith2023-nt, Mass2024-zq}. Its lack of a close modern analogue also makes it a valuable case study for data-driven extreme-event sampling methods \cite{Mahesh2025-py, Whittaker2026-fe}.

Our 66-hour final-conditioned trajectories for the 2021 PNW heatwave begin on June 25, 2021 at 6 PM UTC (11 AM PDT) and end on June 28, 2021 at 12 PM UTC (5 AM PDT). In Figure \ref{fig:free_end_states_pnw_heatwave}, we show mean 2-meter temperature ($T_{\mathrm{2m}}$) over a $126^{\circ}\mathrm{W}$--$110^{\circ}\mathrm{W}$, $38^{\circ}\mathrm{N}$--$55^{\circ}\mathrm{N}$ impact domain with a land mask across 1000-member ensembles for both final- and start-conditioned trajectories in the aforementioned time interval. Rather than reproducing the rapid warming trend observed in the ERA5 ground truth, the final-conditioned members (Figure \ref{fig:free_end_states_pnw_heatwave}q) are, without exception, warmer than ERA5 at the free-end (initial) frame and hold roughly steady while ERA5 strengthens past them into the June 27 peak. The simulated anomalous warmth therefore manifests as an overly intense initial state rather than a building heatwave, with the warmest members exceeding ERA5 by up to +5.8 K over land (Figure \ref{fig:free_end_states_pnw_heatwave}f--h). Similarly, the start-conditioned ensemble members underestimate the day-over-day rise in daily maxima seen in the ERA5 trajectory. Instead, some ensemble members exhibit a dampened diurnal cycle, underestimating both daytime heating and nighttime cooling over land.

\begin{figure}
    \centering
    \includegraphics[width=\textwidth]{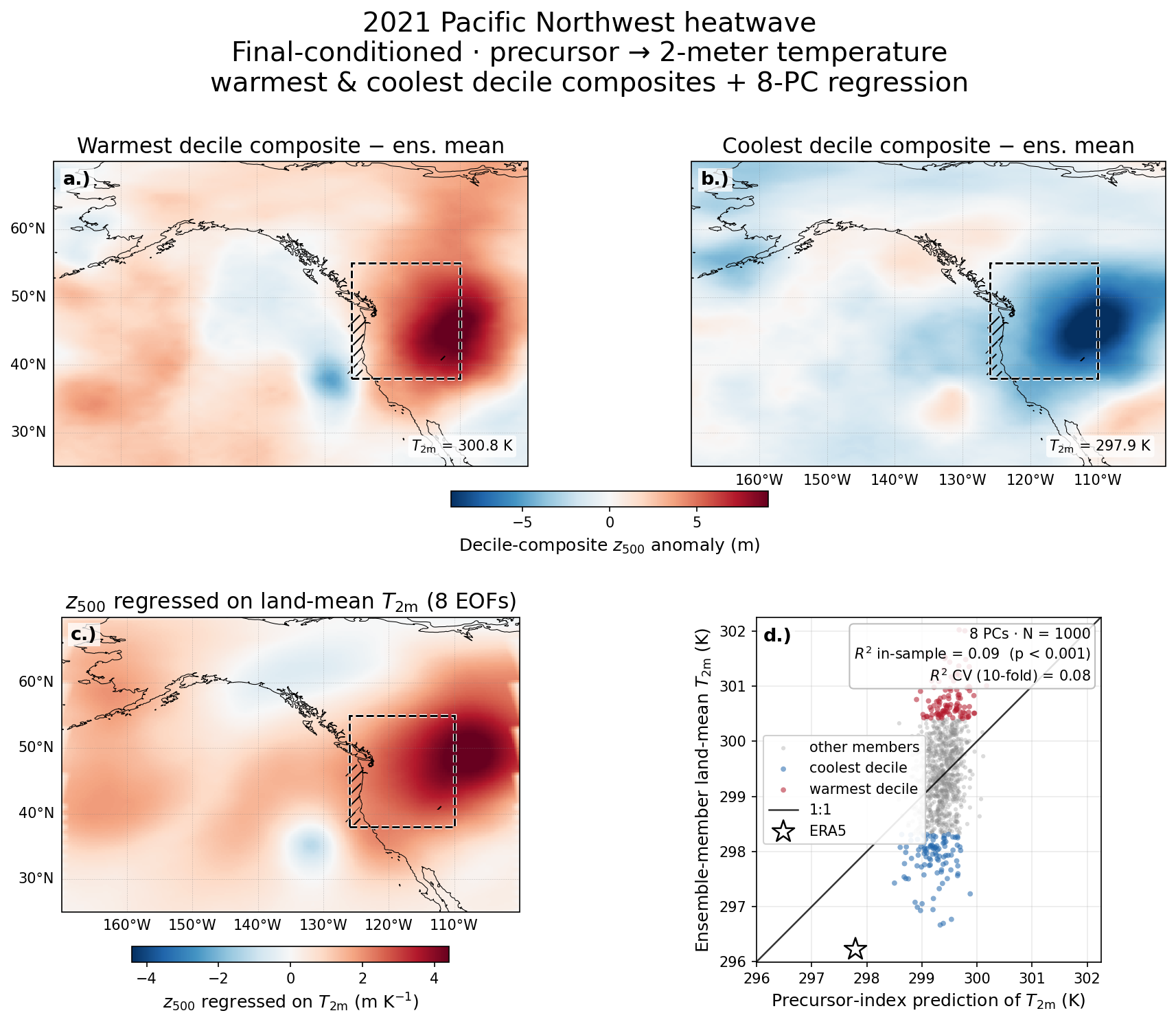}
 \setlength{\belowcaptionskip}{-1em}%

\caption{Precursor $z_{500}$ and land-mean $T_{\mathrm{2m}}$ relationships at the free-end frame \mbox{($t=0$)} for the 2021 PNW heatwave under final-conditioning, with EOF analysis conducted over the $170^{\circ}\mathrm{W}$--$100^{\circ}\mathrm{W}$, $25^{\circ}\mathrm{N}$--$70^{\circ}\mathrm{N}$ domain. (Top) Composite $z_{500}$ anomalies for the warmest (left) and coolest (right) deciles relative to the ensemble mean; dashed boxes indicate the region over which $T_{\mathrm{2m}}$ is averaged ($126^{\circ}\mathrm{W}$--$110^{\circ}\mathrm{W}$, $38^{\circ}\mathrm{N}$--$55^{\circ}\mathrm{N}$). (Bottom left) Slope of spatial regression of EOF-reconstructed $z_{500}$ anomaly onto land-mean $T_{\mathrm{2m}}$ anomaly ($\mathrm{m~K^{-1}}$). (Bottom right) Regressed precursor-index predictions vs. ensemble-member absolute land-mean $T_{\mathrm{2m}}$, highlighting the warmest (red) and coolest (blue) deciles alongside the ERA5 verification (star), 1:1 line, and regression statistics. The star is filled if ERA5's $z_{500}$ state lies within the members' ellipsoid and hollow otherwise. The ellipsoid lives in the space of the leading eight PC scores, is centered on the ensemble mean, is shaped by the members' covariance, and is just large enough to contain every member (Text~S2).}

 \label{fig:eof_regression_pnw_heatwave}
\end{figure}

To quantify the role of large-scale synoptic patterns in the elevated surface temperatures seen in the final-conditioned ensemble members, we first compare the 500 hPa geopotential height composite anomalies over the $170^{\circ}\mathrm{W}$--$100^{\circ}\mathrm{W}$, $25^{\circ}\mathrm{N}$--$70^{\circ}\mathrm{N}$ synoptic domain using the warmest and coolest (by impact domain) deciles, shown in Figure \ref{fig:eof_regression_pnw_heatwave}. We then conduct an empirical orthogonal function (EOF) analysis of 500 hPa geopotential height anomalies on the synoptic domain across all members and perform two regressions to identify the relationship between the EOF-reconstructed geopotential height and the land-mean surface temperature. The first regression, represented by Equation \ref{eq:eof_regression_1_pnw_heatwave}, is a spatial regression of the EOF-reconstructed geopotential height anomaly ($\tilde{z}'_{500}(\mathbf{r})$ at location $\mathbf{r}$ in the synoptic domain) onto the land-mean surface temperature anomaly ($T'_{\mathrm{2m}}$). The slope $b(\mathbf{r})$, representing the change in the reconstructed $z_{500}$ anomaly field per unit change in the land-mean temperature anomaly ($T'_{\mathrm{2m}}$), is shown in Figure \ref{fig:eof_regression_pnw_heatwave}c. The second regression, represented in Equation \ref{eq:eof_regression_2_pnw_heatwave}, is a multiple linear regression that regresses the surface temperature ($T_{\mathrm{2m}}$) against the leading eight principal components (PCs) of the geopotential height precursor field. Figure \ref{fig:eof_regression_pnw_heatwave}d shows the relationship between the multiple linear regression prediction and the ensemble-member surface temperature. We provide further details in Text S2 regarding the EOF and PC conventions, the cross-validation strategy behind reported $R^2$ values, and the formula by which $b(\mathbf{r})$ is computed from the leading eight modes.

\begin{equation}
\label{eq:eof_regression_1_pnw_heatwave}
    \tilde{z}'_{500}(\mathbf{r}) \approx b(\mathbf{r}) T'_{\mathrm{2m}}
\end{equation}

\begin{equation}
\label{eq:eof_regression_2_pnw_heatwave}
    \hat{T}_{\mathrm{2m}} = \beta_0 + \sum_{n=1}^{8} \beta_n \text{PC}_{n}
\end{equation}

The decile-composite anomalies (Figure \ref{fig:eof_regression_pnw_heatwave}a,b) and the spatial regression slope $b(\mathbf{r})$ (Figure \ref{fig:eof_regression_pnw_heatwave}c) reveal a physically consistent relationship: the warmest members at the free end correspond to a positive $z_{500}$ ridge anomaly centered over the Pacific Northwest, while the coolest members feature a trough anomaly. However, this precursor variation is remarkably small: even between the temperature extremes, the warmest- and coolest-decile composites peak at only ${\sim}{\pm}10\mathrm{~m}$ relative to the ensemble mean (the plotted color range, set to the anomaly field's 99th percentile, saturates below these extrema), so the large-scale height pattern barely distinguishes warmer from cooler members. Consistent with this, the leading eight $z_{500}$ PCs of the final-conditioned ensemble explain only 9\% of the across-member variance in land-mean $T_{\mathrm{2m}}$ in-sample (8\% under 10-fold cross-validation; Figure \ref{fig:eof_regression_pnw_heatwave}d)---a weak but statistically robust link---compressing the observed $\sim4$ K spread into a $\sim1$ K predicted range. When conducting this regression on the start-conditioned ensemble, for which the free-end frame is the forecast outcome rather than the antecedent precursor, we also see a comparably weak link between the large-scale height field and the heatwave's average temperature over the impact domain. The leading eight $z_{500}$ PCs explain 8\% of the across-member variance in free-end land-mean $T_{\mathrm{2m}}$ (6\% under 10-fold cross-validation), indicating near-symmetry in explained surface temperature variance between the two conditioning directions. These same eight modes capture 62\% of the across-member variance of the free-end $z_{500}$ field under final-conditioning and 61\% under start-conditioning, although this does not establish that omitted modes are unimportant for explaining temperature variance.

A plausible interpretation of this weak coupling lies in the persistence of the blocking ridge.
Because the heat dome is a quasi-stationary, long-lived feature, its large-scale structure persists across the 66-hour interval, so the amplified ridge enters each member's free-end $z_{500}$ field largely as a shared ingredient pinned by the end state.
Only the residual member-to-member modulation of that shared ridge remains to discriminate warmer members from cooler ones---a modulation whose sign is consistent with the expected ridge--temperature relationship (Figure \ref{fig:eof_regression_pnw_heatwave}a--c) but which explains under 10\% of the temperature spread (Figure \ref{fig:eof_regression_pnw_heatwave}d).
Consequently, while the synoptic ridge is a necessary ingredient of the heat dome, most of the cross-member variance in initial land-mean temperature remains unexplained by linear regression on the leading eight regional height PCs at $t=0$, possibly driven instead by learned variability stemming from smaller-scale thermal perturbations, local boundary layer dynamics, or other non-synoptic variables that remain unconstrained by the end state. Importantly, such processes like land feedbacks are not faithfully represented in the model due to missing variables like soil moisture, and it is possible that this weak statistical relationship reflects model shortcomings rather than learned physical phenomena \cite{Duan2024-jk, Rucker2026-xo}.

\subsection{Superstorm Sandy}

\begin{figure}
    \centering
    \includegraphics[width=\textwidth]{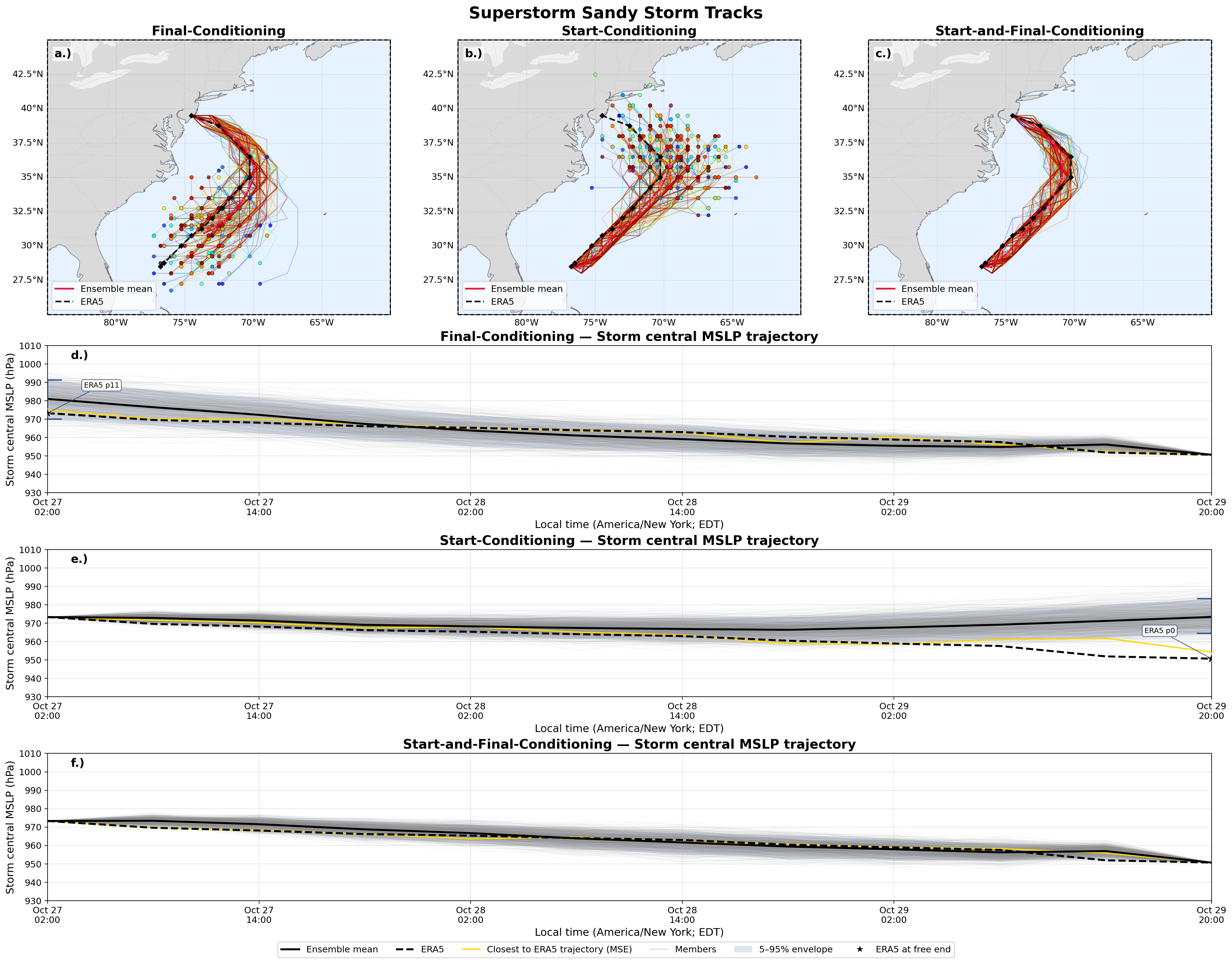}
 \setlength{\belowcaptionskip}{-1em}%

\caption{Superstorm Sandy track and intensity trajectories for the 1000-member ensembles generated across three conditioning modes (start, final, and start-and-final). Hurricane tracks are mapped using TempestExtremes \cite{Ullrich2021-nm}. In all plots, the ERA5 trajectory is plotted as a dashed line, and in subplots (d--f) the ensemble member whose central-pressure trajectory is closest to ERA5's (in mean-squared error) is shown in gold. Blue tick marks at the free-end edge of subplots (d) and (e) denote the ensemble's 5th and 95th intensity percentiles, and the annotated percentile is that of the ERA5 value among the analyzable members.}

 \label{fig:sandy_tc_tracks}
\end{figure}

\begin{figure}
    \centering
    \includegraphics[width=\textwidth]{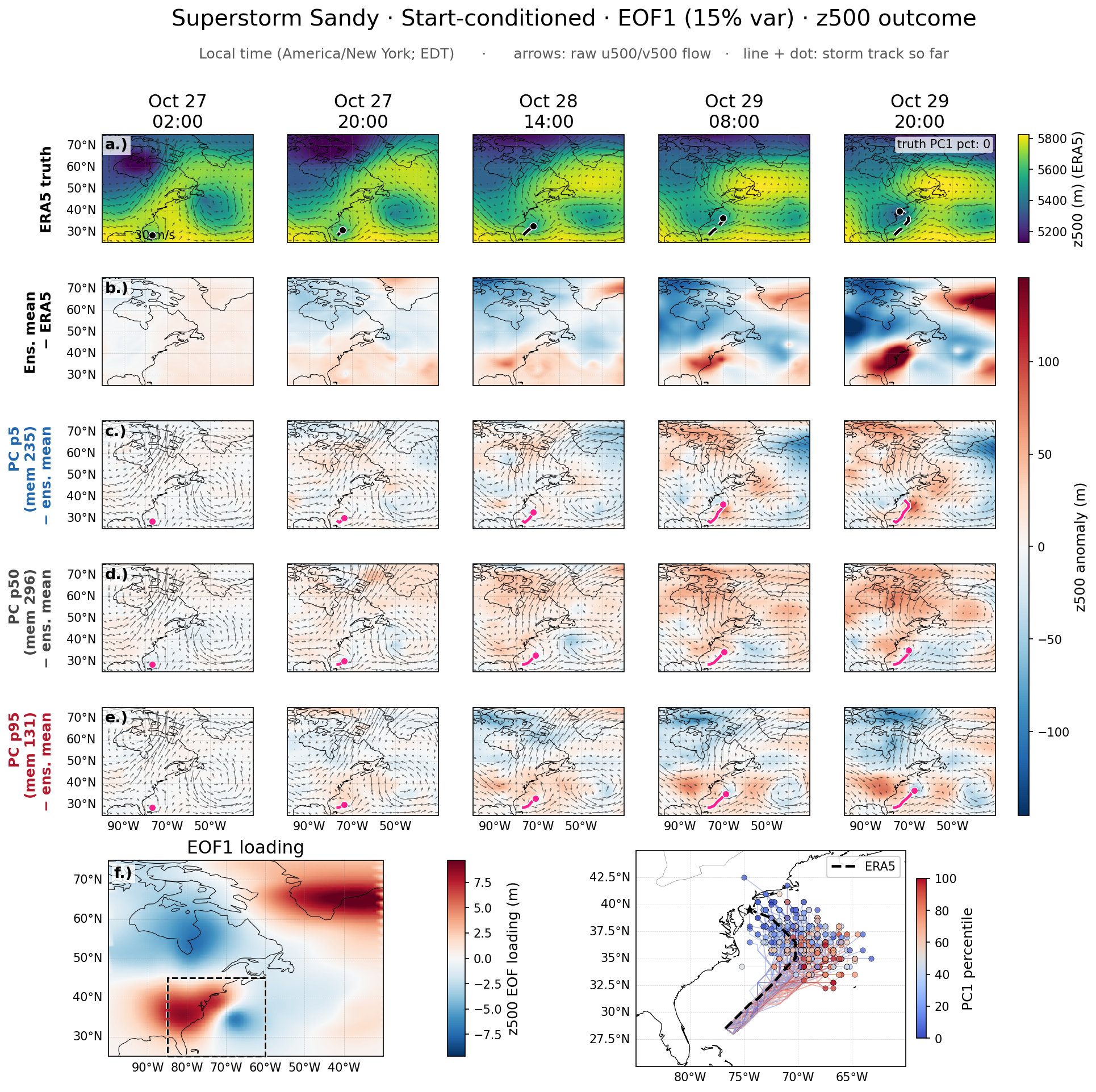}
 \setlength{\belowcaptionskip}{-1em}%

\caption{EOF analysis of the Superstorm Sandy case for start-conditioned ensembles on the 500 hPa geopotential height ($z_{500}$) field over the domain $100^{\circ}$--$30^{\circ}$~W, $25^{\circ}$--$75^{\circ}$~N at the free-end frame (October 30, 2012, 12 AM UTC). The first row shows the evolution of ERA5 $z_{500}$, the second row the start-conditioned ensemble-mean difference from ERA5, and the third through fifth rows $z_{500}$ anomalies relative to the ensemble mean for individual members at the 5th, 50th, and 95th percentiles of the first EOF. The final row shows the EOF1 loading and hurricane tracks colored by each member's EOF1 percentile; ERA5 sits at the 0th percentile of this mode. The dashed box in the loading panel shows the track-plotting domain ($85^{\circ}$--$60^{\circ}$~W, $25^{\circ}$--$45^{\circ}$~N).}

 \label{fig:sandy_start_eof1}
\end{figure}

In October of 2012, Superstorm Sandy struck the Eastern Seaboard nearly perpendicularly, having made an anomalous westward turn into the New Jersey coast that claimed over a hundred lives and caused billions of dollars in damages \cite{Centers-for-Disease-Control-and-Prevention-CDC-2013-ua, Hall2013-yl, Barnes2013-np}. This trajectory, steered by a shift in the mean winds and a blocking event caused by cyclonic wave-breaking, lacked precedent in the historical record and is estimated to have a $\sim$700-year return period based on a 50,000-year simulation \cite{Hall2013-yl, Barnes2013-np}. Given the unique circumstances that facilitated Sandy's anomalous track and intensification, better understanding the governing dynamics that could transform otherwise ordinary storms into similar gray swans is of high scientific and societal importance \cite{Galarneau2013-ff, Hakim2026-oy}. Most recently, \citeA{Hakim2026-oy} optimized initial conditions for five ``ordinary'' hurricanes with northward tracks over the western Atlantic at a similar time of year to also make landfall on the coast. While their results indicate that highly specific perturbations to an ordinary hurricane can yield Sandy-like outcomes, a 1000-member ensemble experiment that randomly perturbed initial conditions for one of these hurricanes (Fiona) did not yield any similar outcomes. This motivates our approach of conditioning at the end, as small perturbations to one set of initial conditions cannot fully sample the diversity in starting conditions that ultimately end in Sandy-like landfall.

For our 1000-member ensembles, we choose the 66-hour interval starting on October 27, 2012 at 6 AM UTC (2 AM EDT) and ending on October 30, 2012 at 12 AM UTC (October 29, 8 PM EDT), spanning Sandy's approach and New Jersey landfall. We map the tracks and intensity trajectories (central MSLP) of these ensemble members alongside the ERA5 ground truth in Figure \ref{fig:sandy_tc_tracks}. For free-end statistics, we analyze only members whose storm is detected by the tracker at the free-end frame, hereafter referred to as ``analyzable members.'' This leaves us with 788 of 1000 final-conditioned members and 984 of 1000 start-conditioned members (Text S3). In the final-conditioned ensemble, members pinned to Sandy's historical landfall state at $t=66$~h exhibit large diversity in initial hurricane positions at $t=0$~h (Figure \ref{fig:sandy_tc_tracks}a), with member storm centers beginning at a wide range of locations across the western Atlantic and lying a median of ${\sim}440$~km from the observed starting position. They also exhibit a broad range in intensity, with MSLP ranging from a minimum of 958~hPa to a maximum of 997~hPa (Figure \ref{fig:sandy_tc_tracks}d). Moreover, the final-conditioned tracks show no evidence of discrete clustering, suggesting that the model's generated distribution of antecedent states consistent with Sandy's landfall forms a smooth continuum that final-conditioning can traverse directly---a sampling that may be intractable using initial-condition perturbation. In terms of storm-center positional spread, this antecedent diversity is comparable in magnitude to the corresponding free-end spread of the start-conditioned ensemble (Figure \ref{fig:sandy_tc_tracks}b). At the free-end (final) frame of the interval for the start-conditioned ensemble, member storm centers lie a median of ${\sim}590$~km from the observed landfall fix, and 99.3\% of them sit east of it. Because the interval closes at the true landfall hour, members offshore at the free end may still strike the coast at later times, so we characterize track diversity at the common free-end time rather than by eventual landfall. Start-and-final-conditioning, by contrast, restricts tracks to a narrow corridor interpolating between the two pinned states (Figure \ref{fig:sandy_tc_tracks}c,f). Surprisingly, while the intensities of the start-conditioned hurricanes span a wide range (954 to 993~hPa; Figure \ref{fig:sandy_tc_tracks}e), none of them reproduce the rapid intensification seen with Superstorm Sandy: no analyzable member is as deep as Sandy's observed 950.7~hPa at the free-end (final) frame.

To quantify how strongly the large-scale flow organizes the diversity at the free end for both final- and start-conditioned ensembles, we perform an EOF analysis of the free-end $z_{500}$ field over the domain $100^{\circ}$--$30^{\circ}$~W, $25^{\circ}$--$75^{\circ}$~N and regress three per-member storm properties at the free-end frame---latitude, longitude, and central MSLP, identified from each member's resolved storm track---on the leading eight principal components, exactly paralleling Equation \ref{eq:eof_regression_2_pnw_heatwave}. Unlike the 2021 PNW heatwave case, however, the regressions use only the analyzable members defined above. All reported $R^2$ values are deterministic 10-fold cross-validated values. We provide additional details regarding the tracker settings in Text S3.

Under final-conditioning, where the free end is the precursor, the leading eight PCs explain 48\% (21\%) of the cross-member variance in antecedent storm latitude (longitude) and roughly 15\% in intensity (Figure~S2). Because predictors and targets are evaluated at the same free-end frame, these regressions measure statistical organization rather than establish steering causality: storm-position differences can also contribute to variability in $z_{500}$. Furthermore, no single synoptic mode carries the final-conditioned organization. EOF2 (11.6\% of the $z_{500}$ field's variance) co-varies with the antecedent track only weakly, and this is not an artifact of the coordinate axes: no orientation we tested increases the Spearman rank correlation above $\rho \approx -0.23$, barely stronger than the $-0.22$ value obtained when using latitude alone. ERA5 itself sits at the 4th percentile of PC2. The PC most strongly rank-correlated with antecedent storm position (PC6, $\rho \approx -0.42$) has its loading (EOF6) extremum on the storm itself, consistent with a contribution from storm-position differences rather than solely from environmental variability. The next strongest (PC7, PC8) may likewise partly reflect storm-position differences. The eight-PC regression therefore explains 48\% of the variance in antecedent storm latitude, but this association cannot be attributed solely to environmental steering because the height-field variability may also reflect differences in storm position.

Under start-conditioning, by contrast, the large-scale height field organizes this diversity far more strongly. The start-conditioned ensemble marks the strongest observed relationship here: the leading eight PCs explain 79\% of the variance in latitude, 44\% in longitude, and 19\% in MSLP (Figure~S3). The contrast with the 2021 PNW heatwave case holds in both conditioning directions: the same eight-PC regression explains at most 8\% of the temperature diversity, but 48\% of Sandy's antecedent storm latitude variance under final-conditioning and 79\% of its final-frame storm latitude variance under start-conditioning. Furthermore, a single mode does emerge under start-conditioning to organize this spread. EOF1 (15\% of the $z_{500}$ field's variance) shown in Figure \ref{fig:sandy_start_eof1} functions as a landfall-proximity axis: members in its lowest percentiles end the interval at the coast near Sandy's observed position while members in its highest percentiles end far offshore to the southeast. While at least one ensemble member ends the 66-hour sequence further inland than Superstorm Sandy, the ERA5 $z_{500}$ field and Sandy's observed central pressure lie at the 0th percentile of PC1 and free-end MSLP, respectively. Across the leading eight PCs, ERA5's $z_{500}$ state lies farther from the ensemble mean than any member's (Text~S2). The failure to sample Sandy's observed free-end state is qualitatively consistent with previous work documenting the difficulty of sampling Sandy-like trajectories using conventional forward ensembles \cite{Hall2013-yl, Hakim2026-oy}. However, this failure likely also reflects model shortcomings or finite-ensemble sampling and does not, by itself, establish event rarity as the explanation.

\subsection{Hurricane Ian}

\begin{figure}
    \centering
    \includegraphics[width=\textwidth]{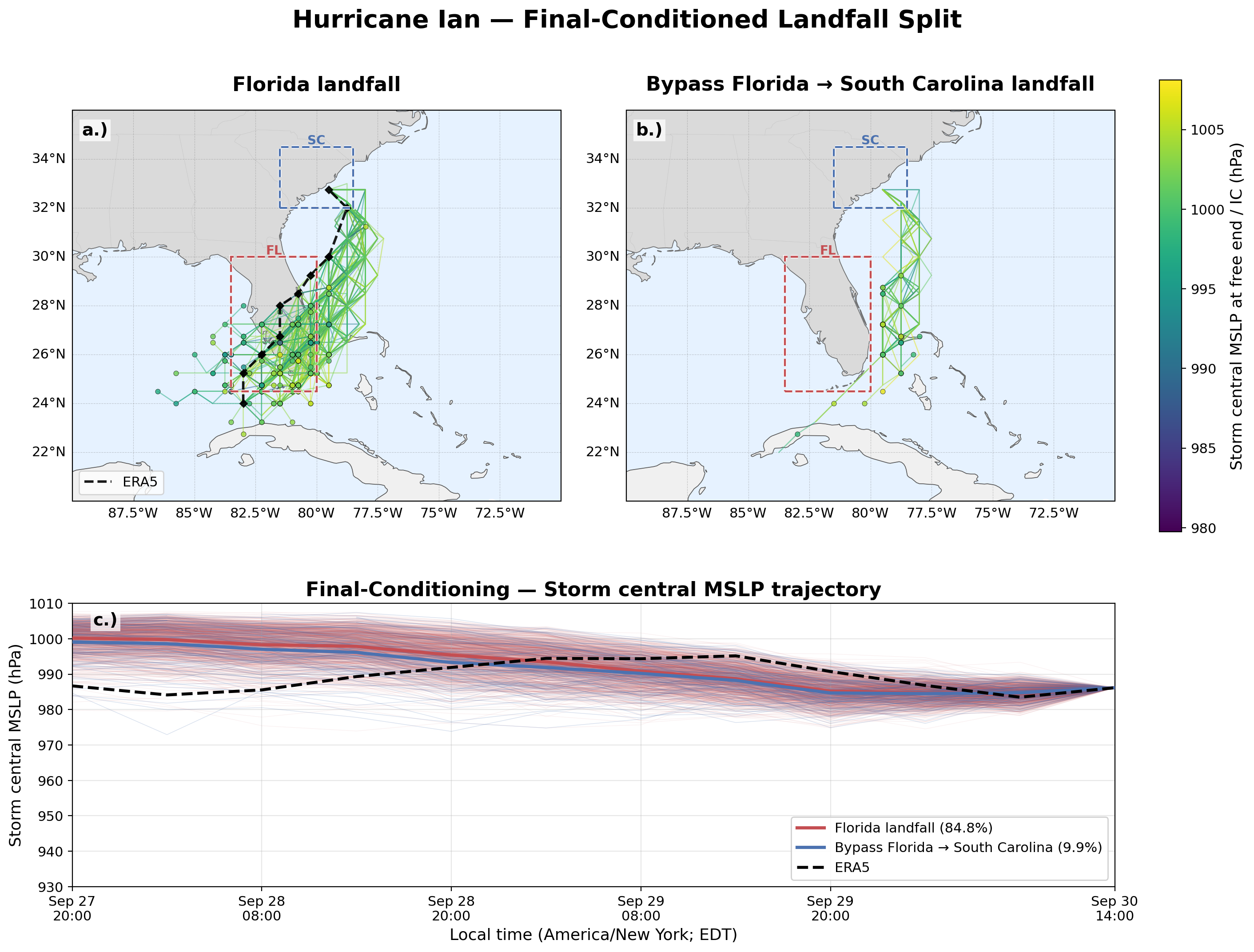}
 \setlength{\belowcaptionskip}{-1em}%
\caption{Hurricane Ian track and intensity trajectories for the final-conditioned ensemble split across two cases---making initial landfall in Florida vs. bypassing Florida entirely.
Tracks connect 6-hourly TempestExtremes storm-center fixes \cite{Ullrich2021-nm}, effectively resolved at the model's native HPX64 ($\sim 100$~km) scale (Section 2); the angularity is a discrete-sampling artifact, most pronounced where the central-pressure minimum is weak or diffuse (e.g., near genesis), and does not reflect physical track variability.
Hurricane Ian's actual trajectory and intensity are plotted as dashed lines in subplots a and c.}
 \label{fig:ian_landfall_split}
\end{figure}

\begin{figure}
    \centering
    \includegraphics[width=\textwidth]{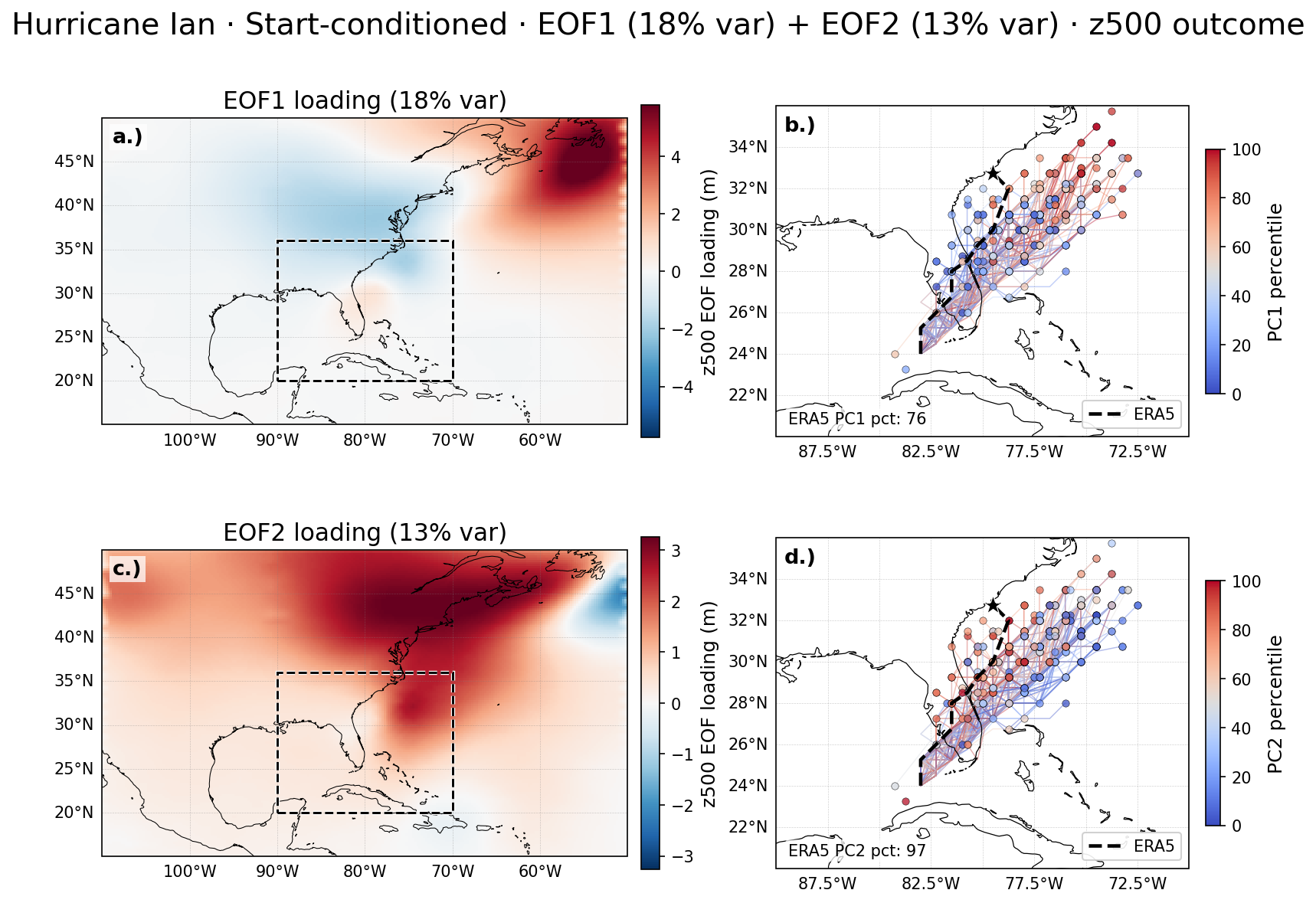}
 \setlength{\belowcaptionskip}{-1em}%
\caption{The two leading start-conditioned $z_{500}$ EOFs for the Hurricane Ian case, computed at the free-end frame (September 30, 2022, 18:00 UTC) over the domain $110^{\circ}$--$50^{\circ}$~W, $15^{\circ}$--$50^{\circ}$~N.
Each row pairs one EOF's loading pattern (left; the dashed box marks the track domain shown at right) with the member storm tracks and free-end fixes over the track domain ($90^{\circ}$--$70^{\circ}$~W, $20^{\circ}$--$36^{\circ}$~N), colored by each member's percentile on that row's PC score; the dashed black line is the ERA5 track and the star its free-end fix.
EOF1 (a, b; 18.4\% of the $z_{500}$ field's variance) stays weak over the storm itself, and its PC sorts members along the coast; ERA5 sits at its interior 76th percentile.
EOF2 (c, d; 13.3\% of the $z_{500}$ field's variance) is a broad, single-signed ridge extending far beyond the storm, and its PC sorts members shoreward, toward the observed fix; ERA5 sits at its 96.6th percentile.}
 \label{fig:ian_start_eof_tracks}
\end{figure}

In September 2022, Hurricane Ian devastated the southeastern United States, striking Florida as a Category 4 storm and South Carolina as a Category 1. With over \$112 billion in damages and more than 150 fatalities, it ranks as the costliest hurricane in Florida's history and the third costliest in U.S. history \cite{bucci2023hurricaneian, Hauptman2024-xt}. Because Ian made multiple U.S. landfalls, it serves as a prime example of a single system inflicting sequential damage across distinct regions. For final-conditioned ensembles, such events highlight both the variation in dynamical precursors and the specific upstream regions placed at risk. Consequently, Hurricane Ian provides a hurricane case in which antecedent hurricane characteristics discovered by final-conditioning (i.e. storm position and intensity at the starting frame) map to differences in regional hazard risk.

To examine this diversity and assess how strongly antecedent storm position and intensity are statistically organized by the leading regional $z_{500}$ modes, we analyze a 66-hour interval spanning Ian's approach, Florida landfall, and final South Carolina landfall, starting on September 28, 2022 at 00:00 UTC (September 27, 8:00 PM EDT) and ending on September 30, 2022 at 18:00 UTC (September 30, 2:00 PM EDT). Tracks and central-pressure trajectories for all three conditioning modes are shown in Figure~S1. Across the final-conditioned ensemble members, $84.8\%$ (848 members) make landfall in Florida (Figure \ref{fig:ian_landfall_split}a) and a small minority ($9.9\%$, 99 members) bypass Florida entirely, traveling north along the Atlantic coast (Figure \ref{fig:ian_landfall_split}b). The remaining $5.3\%$ (53 members) do not have a storm that the tracker detects at the first frame and are not analyzed. The Florida-landfall criterion and the track quality control behind these counts are given in Text S3. Compared with Sandy, Ian's antecedent track diversity is more weakly organized by the large-scale height field, as measured by our eight-PC linear regressions. While the eight leading PC modes capture 73\% of the antecedent $z_{500}$ field's variance across all 1000 members, the EOF regression analysis on the 947 analyzable members explains no more than $7\%$ of variance in antecedent latitude and longitude and only $12\%$ in intensity.

Under start-conditioning---where the ensemble is pinned to Ian's initial state in the Gulf of Mexico and the 66-hour endpoint is left free---the large-scale height field organizes the free-end track far more strongly than under final-conditioning, consistent with the behavior observed for Sandy. The leading eight principal components, computed over the $110^{\circ}$--$50^{\circ}$~W, $15^{\circ}$--$50^{\circ}$~N domain, explain 52\% of the cross-member variance in free-end longitude (comparable to Sandy's start-mode 44\%) and 40\% in latitude (Figure~S5). Both are far more than the $\sim$7\% organized under final-conditioning (Figure~S4), indicating a sharper start-versus-end asymmetry than that of Sandy. As with Sandy, the two leading modes have broad spatial structure extending beyond the storm, consistent with an environmental contribution to these statistical associations. Furthermore, the two leading modes separate cleanly: EOF1 organizes how far along the coast an ensemble member ends the interval, while EOF2 organizes its final-frame distance from the coast. Along the EOF1 axis, which explains $\sim$18.4\% of the $z_{500}$ field's variance (Figure \ref{fig:ian_start_eof_tracks}a,b), PC1 sorts free-end latitude and longitude at $\rho = 0.45$ and $0.44$, respectively, and Ian sits in the middle of the ensemble while the corresponding ERA5 $z_{500}$ height field falls at the 76th percentile of PC1. Along the EOF2 axis, which explains 13.3\% of the $z_{500}$ field's variance (Figure \ref{fig:ian_start_eof_tracks}c,d), PC2 sorts free-end longitude at $\rho = -0.33$. Here, Ian sits ${\sim}300$~km shoreward of the ensemble's mean free-end position---west of 82.8\% of member fixes---and the ERA5 $z_{500}$ state falls at the 96.6th percentile of PC2. This is the same edge geometry seen in Sandy's start-conditioned ensemble, where ERA5 sits at the 0th percentile of the mode that organizes landfall proximity. In terms of intensity, Ian also sits at the tail of both start- and final-conditioned ensembles, with an intensity exceeding 98\% of the analyzable members in both conditioning directions (Figure~S1, Figure \ref{fig:ian_landfall_split}c). Finally, we note that under start-conditioning, the tracker does not detect the storm at the final frame in 243 of the 1000 members (Text S3), so these free-end statistics describe the remaining 757.

\section{Discussion}

Generative video diffusion climate emulators like cBottle-video allow one to ask questions that are impractical for traditional physics-based and data-driven autoregressive models. Those approaches, though methodologically distinct, address the same fundamental question (variational assimilation and adjoint sensitivity being partial exceptions): what trajectories are possible given an initial, instantaneous state of the atmosphere? The question we seek to answer instead is: given a realized extreme, is there a diverse set of synoptic histories that could have produced it? Our experiments suggest that this may be the case. For Hurricanes Sandy and Ian, generated hurricane tracks over the 66 hour window are nearly as diverse backward as forward. Trajectories conditioned on Hurricane Ian making landfall in South Carolina suggest material consequences: varying tracks make prior landfall in very different parts of Florida, or skip Florida altogether. For the 2021 PNW heatwave, alternative histories consistent with the same end state show an earlier onset with persistently elevated temperatures, rather than a rapid multi-day intensification. Understanding these generated histories, however, requires examining where this antecedent diversity lives and how it is organized.

Across all three cases, the geopotential height field does not organize the majority of the antecedent diversity: the leading eight modes explain 48\% of hurricane track latitude for Sandy and $\sim$6\% of hurricane track longitude for Ian (both among analyzable members), and 8\% of the impact-domain land-mean $T_{\mathrm{2m}}$ for the 2021 PNW heatwave. For the hurricanes, this exposes a marked asymmetry---the height field organizes free-end storm position far more strongly when that free end is the forecast outcome (79\% for Sandy's latitude, 52\% for Ian's longitude) than when it is the antecedent. By contrast, the heatwave exhibits near-symmetry with weak organization in both directions (8\% backward vs. 6\% forward cross-validated), bounding the asymmetry to the storm cases while demonstrating that low explanatory power is not an inherent artifact of final-conditioning. Final-conditioned ensembles retain 84--89\% of the start-conditioned $z_{500}$ spread, demonstrating substantial height-field variability. The leading eight modes capture 62\% and 73\% of the final-conditioned free-end height variance for the heatwave and Ian, respectively. However, capturing substantial height variance does not exclude predictive relationships involving omitted modes or nonlinear structure. Storm-position signatures in some final-conditioned modes also complicate interpreting the regression scores as measures of environmental influence. What organization exists is distributed across several modes rather than a single steering pattern, and, as Sandy's antecedent tracks show, alternate histories form a smooth continuum rather than discrete synoptic clusters. Because a continuous, weakly organized precursor space cannot be enumerated by a catalog of weather regimes, this motivates directly sampling the trajectory distribution---a capability that arbitrary conditioning unlocks but one that forces the question of whether every generated trajectory is physically admissible.

Because cBottle-video is trained using a denoising score-matching objective without hard constraints to enforce physical laws, the pretrained checkpoint used here is capable of generating a sequence conditioned on any pair of atmospheric states, lacking an intrinsic filter to prevent generation of dynamically inadmissible trajectories. To learn the extent to which cBottle-video can interpolate between chronologically disconnected atmospheric states, we construct three deliberately contrived conditioning stress tests shown in Figures~S12--S14: interpolating between tropical cyclones on opposite sides of the equator (Cyclone Agni at \ang{0.5}N on 27 November 2004, 12:00 UTC and Tropical Storm Abaimba at \ang{5.8}S on 2 October 2003, 12:00 UTC), interpolating between swapped endpoints for Superstorm Sandy (i.e. starting at Sandy's landfall on 30 October 2012, 00:00 UTC and ending 66 hours prior on 27 October 2012, 06:00 UTC), and interpolating between the August 2011 Texas heatwave and the February 2021 Texas freeze (specifically 28 August 2011, 18:00 UTC and 15 February 2021, 12:00 UTC). For all three cases, the calendar embedding and interpolated monthly SST assume a 66-hour sequence consistent with the timestamp used for the start frame. In the example interpolating between Cyclone Agni in the Northern Hemisphere and Tropical Storm Abaimba in the Southern Hemisphere, the model smoothly dissolves the Northern Hemisphere low over four frames, passes through an unorganized intermediate state, and nucleates a separate Southern Hemisphere cyclone to satisfy the terminal constraint. In the example interpolating between swapped endpoints for Superstorm Sandy, the storm making landfall in the start frame dissipates and a new storm forms just south of Sandy's position 66 hours before landfall and drifts north as it deepens into the final frame. In both cases, the model treats the two cyclones at the start and final frames as separate systems, dissolving one and generating the other, so each step looks locally plausible even though no physical 66-hour trajectory connects the two global states. The final case allows no such local workaround: the model must take the contiguous United States from summer to winter within 66 hours. In our non-contrived case studies, the model is never forced to bridge incompatible states. Conditioning is applied to single realized endpoints when conditioning on a single frame and to opposite ends of a 66-hour ERA5 window when interpolating between two atmospheric states. Yet because the emulator will interpolate across contrived state pairs without signaling unphysicality, dynamical plausibility cannot be taken for granted by construction. Instead, methods for explicit verification must be developed before final-conditioned ensembles generated using video diffusion models like cBottle-video can be reliably used for scientific discovery and scenario planning. These efforts are ongoing.

For final-conditioned ensembles, the realized past is already known, so verification metrics designed to reward alignment with a single historical trajectory are insufficient on their own for verifying physically plausible counterfactual diversity. However, physical balance checks and comparative ensemble statistics can expose unique failure modes and structural biases. To this end, we evaluate the 500~hPa midlatitude (\ang{30}-\ang{60} N and \ang{30}-\ang{60} S) ageostrophic fraction and the ratio of ensemble spread to ensemble-mean root-mean-square error (RMSE) across final-, start-, and start-and-final-conditioned ensembles. Across all three cases, the magnitude of the 500 hPa midlatitude winds systematically weakens further away from the pinned end of the generated ensemble, as seen in the second row of Figures~S15--S17. While the magnitude of the implied geostrophic wind is largely consistent near the pinned frames, the ageostrophic fraction of the total wind rises in tandem, as seen in the top row of the same set of figures. In the case of Superstorm Sandy, this rise in the ageostrophic fraction is clearly attributable to an increase in the magnitude of the ageostrophic wind, as seen in the last row of Figure S16. But for Hurricane Ian and the 2021 Pacific Northwest heatwave, the magnitude of the geostrophic and ageostrophic wind stays relatively steady close to the pinned end while the weakening of the total wind is immediate, indicating that the weakening of the total wind reflects an ageostrophic component that opposes the geostrophic flow, leaving the wind more subgeostrophic than the height field implies. This surfaces a physical inconsistency between the 500 hPa geopotential and the actual winds at this pressure level. When it comes to the ratio between ensemble spread and ensemble-mean RMSE (i.e. spread-skill ratio) for 500 hPa geopotential height in the regional case-study domains used to calculate the leading $z_{500}$ modes, the value never exceeds 0.9 (and is generally much lower) across all conditioning modes, indicating structural underdispersion and ensemble-mean bias. In line with this, ERA5's antecedent $z_{500}$ state lies outside the final-conditioned members' ellipsoid in every case (Text~S2; Figures~2, S2, and S4). While future iterations of video diffusion climate emulators like cBottle-video will likely mitigate the issues presented thus far, we believe that future work should make use of a more comprehensive suite of dynamical, spectral, and conservation metrics \cite{Kelder2022-xb, Hakim2024-dq, Kasteleyn2026-jf}. Ultimately, final validation prior to depending on this method will likely require verifying that forward integrations from the discovered initial conditions converge towards the observed end state in a physics-based model.

Such physical inconsistencies reflect the limitations of evaluating an off-the-shelf video diffusion checkpoint that was not fine-tuned for this task, rather than an intrinsic failure of generative final-conditioning. Future iterations of these models that incorporate targeted interventions for the aforementioned deficiencies may already be possible given emerging research. For example, in \citeA{Baldan2025-de}, the authors integrate physical constraints into flow matching models (a close cousin of diffusion models) using a combination of techniques that result in distributional and physical accuracy. Although cBottle-video is a video diffusion model, flow matching models have already been extended to video generation with promising results \cite{Jin2024-wd}. Furthermore, generative models that are inherently bidirectional can enable unique opportunities for self-supervised error estimation, as demonstrated by \citeA{Scheinker2026-xr}. When employed during training, it is possible that this could be used to mitigate compounding degradation resulting from reproducing conditioned frames.

\section{Conclusion}

Preparing for high-impact climate extremes requires understanding their plausible precursor pathways, yet searching for rare events by forward-integrating physics-based ensembles scales prohibitively with lead time and rarity. Here, we address this bottleneck by casting extreme history generation as a final-conditioning problem in a video diffusion climate emulator, targeting realized disaster states directly. Across our historical hurricane and heatwave case studies, final-conditioned ensembles retain 84--89\% of the synoptic spread observed in forward forecasts, demonstrating that reverse-time sampling does not collapse diversity. The large-scale 500 hPa height field organizes only a fraction of this antecedent variance, with alternative histories forming a continuous synoptic spectrum rather than isolated clusters. This is likely due, in part, to the short 66-hour window necessitated by the cBottle-video design. Direct generative sampling using this method suggests that identical terminal extremes may emerge from structurally distinct physical evolutions.

Such final-conditioned trajectories occupy a space unreachable via conventional forecasting frameworks. While novel techniques for identifying plausible stronger extremes---such as initial condition optimization, ensemble boosting, and rare event sampling---are also emerging at a steady pace, none sample the full distribution of plausible histories consistent with a single event \cite{Ragone2018-rj, Gessner2021-qy, Finkel2024-ln, Finkel2026-wn, Whittaker2026-fe, Hakim2026-oy, Lancelin2026-om}. Capturing this distribution is essential because different pathways to the same terminal disaster can produce markedly different local impacts beforehand, as suggested by the divergent earlier landfalls across generated trajectories for Hurricane Ian. If final-conditioning methods prove to be robust, this capability would vastly expand the scope of ``event-based storylines'' needed to convey the escalating risks to lives and property from compound extremes in a warming climate \cite{Shepherd2018-zo, Baldwin2019-ec, Sillmann2021-jl, Keys2023-zx, Whittaker2026-fe}. By pairing final-conditioned sampling with guided likelihood estimation \cite{Manshausen2026-nu}, generative emulators may one day attach calibrated probabilities to these diverse physical narratives, transforming extreme-event discovery into an actionable tool for climate risk assessment.

\section*{Data and Code Availability}

All code used in this study is archived on Zenodo \cite{lin_2026_22000716} and developed openly at \url{https://github.com/jerrylin96/rewinding-the-extremes}. ERA5 reanalysis data \cite{Hersbach2020-oh} used for conditioning and verification was downloaded through WeatherBench 2 \cite{rasp2023weatherbench} using the Earth2Studio software repository \cite{Earth2Studio_Contributors_NVIDIA_Earth2Studio_2024}. Earth2Studio was also used to load the cBottle-video checkpoint from \citeA{Brenowitz2025-mz} and run inference with it. The checkpoint is publicly available at \url{https://huggingface.co/nvidia/cbottle/blob/eebd93c85b3cd3a5a8f79c546ed917b0b80438f4/cBottle-video.zip}. Earth2Studio's inference calls the sampler implemented in the cBottle package (\url{https://github.com/NVlabs/cBottle}). Regridding between the $0.25^{\circ}$ latitude--longitude grid and HPX64, in both directions, used earth2grid (\url{https://github.com/NVlabs/earth2grid}).

\section*{Author Contributions}

Jerry Lin: Software, Conceptualization, Writing -- original draft, Writing -- review \& editing.
Elizabeth A. Barnes: Supervision, Funding acquisition, Conceptualization, Writing -- review \& editing.
Mu-Ting Chien: Writing -- review \& editing.
Mansi Sakarvadia: Writing -- review \& editing.

\section*{Funding}

J.L. was supported by the DOE Office of Science through the Program for Climate Model Diagnosis and Intercomparison (PCMDI). M.C. was supported by Heising-Simons Foundation grant \#2023-4720. M.S. was supported by the U.S. Department of Energy, Office of Science, Office of Advanced Scientific Computing Research, Department of Energy Computational Science
Graduate Fellowship under Award Number DE-SC0023112.

% \section*{Acknowledgements}

\section*{Use of Generative AI and AI-Assisted Technologies}

Claude Opus 5.5 through 4.8 and Fable 5 in Claude Code were used for editing code and text, assisting with literature searches, and drafting the first version of the Supporting Information. GPT-6.1 Sol in Codex was used for adversarial review of git commits. Gemini 3.8 Flash in Antigravity CLI was used to edit text for flow and clarity. Arena.ai was used to iterate on the wording of overly complex sentences. GPT-5.6 Sol in OpenAI Prism was used to fix and adjust LaTeX formatting prior to manuscript submission. The authors critically reviewed and verified all AI-assisted material, tested all AI-assisted code, independently verified all cited sources, and take full responsibility for the accuracy, integrity, and originality of the manuscript.

\section*{Competing Interests}

The authors declare no competing interests.

\bibliography{references}

\clearpage

%%%%%%%%%%%%%%%%%%%%%%%%%%%%%%%%%%%%%%%%%%%%%%%%%%%%%%%%%%%%%%%%%%%%%%%%%%%%
% AGUtmpl.tex: this template file is for articles formatted with LaTeX2e,
% Modified December 2018
%
% Supporting Information for AGU Journals
%%%%%%%%%%%%%%%%%%%%%%%%%%%%%%%%%%%%%%%%%%%%%%%%%%%%%%%%%%%%%%%%%%%%%%%%%%%%

\graphicspath{{figures/}} % committed figures live in paper/figures/ (FIGURES.md)
\setkeys{Gin}{draft=false}
\raggedbottom

% Number Supporting Information equations and figures independently.
\setcounter{equation}{0}
\renewcommand{\theequation}{S\arabic{equation}}
\setcounter{figure}{0}
\renewcommand{\thefigure}{S\arabic{figure}}

%% ------------------------------------------------------------------------ %%
%  TITLE
%% ------------------------------------------------------------------------ %%

\title{Supporting Information for ``Extremes on Rewind: Generating 1,000-Member Ensembles Initialized at a Final Condition''}

%% ------------------------------------------------------------------------ %%
%  AUTHORS AND AFFILIATIONS
%% ------------------------------------------------------------------------ %%

\authors{Jerry Lin$^{1}$, Mu-Ting Chien$^{1}$, Mansi Sakarvadia$^{2}$, Elizabeth A. Barnes$^{1}$}

\affiliation{1}{Department of Computing \& Data Sciences, Boston University, Boston, MA, USA}
\affiliation{2}{Department of Computer Science, University of Chicago, Chicago, IL, USA}

%% ------------------------------------------------------------------------ %%
%  TEXT
%% ------------------------------------------------------------------------ %%

\noindent\textbf{Contents of this file}
\begin{enumerate}
\item Text S1
\item Text S2
\item Text S3
\item Text S4
\item Text S5
\item Figures S1 to S17
\end{enumerate}

\noindent\textbf{Introduction}
This supporting information document provides supporting text (Texts S1 to S5) and auxiliary figures (Figures S1 to S17) accompanying the manuscript ``Extremes on Rewind: Generating 1,000-Member Ensembles Initialized at a Final Condition''. Text S1 details the diffusion sampler, sea surface temperature conditioning, and member seeding. Text S2 defines empirical orthogonal function (EOF) and principal component (PC) conventions, SVD formulation, regressions, and cross-validation. Text S3 documents TempestExtremes storm tracking, candidate filtering, and ensemble quality control. Text S4 defines free-end $z_{500}$ ensemble spread across conditioning modes. Text S5 defines the 500~hPa ageostrophic wind diagnostic and its regridding control. Figure S1 presents Hurricane Ian tracks and central-pressure trajectories across all three conditioning modes. Figures S2 to S5 show free-end track and intensity regressions on leading $z_{500}$ principal components for Superstorm Sandy (S2, S3) and Hurricane Ian (S4, S5). Figures S6 to S11 report lead-time verification metrics (spread, signed mean error, RMSE, CRPS) against ERA5 for $z_{500}$ and surface impact variables across all cases. Figures S12 to S14 present conditioning stress tests using endpoint states that do not form a chronological 66-hour sequence. Figures S15 to S17 report 500~hPa ageostrophic diagnostics for the 2021 Pacific Northwest heatwave (S15), Superstorm Sandy (S16), and Hurricane Ian (S17).

\noindent\textbf{Text S1. Sampling configuration, SST conditioning, and member seeding}

Samples are drawn by integrating Equation 2 of the main text from $\sigma = 1000$ to $\sigma = 0$. The denoiser is evaluated only at eighteen noise levels spaced by the $\rho = 7$ schedule of Karras et al. (2022) between $\sigma_{\max} = 1000.0$ and $\sigma_{\min} = 0.02$, matching the training noise range (Brenowitz et al., 2025, Section A.8.2). Each step between these levels uses a second-order Heun update, and the final step from $\sigma_{\min}$ to $\sigma = 0$ is a single Euler step that returns the denoiser's estimate $\mathcal{D}_\theta(\mathbf{x}_{\sigma_{\min}}; \sigma_{\min}, \mathbf{c})$. No stochastic noise is added at any step ($S_{\text{churn}} = 0$).

Sea surface temperature conditioning uses the AMIP mid-month SST boundary-condition dataset distributed with the checkpoint, with bracketing monthly values linearly interpolated in time to each frame's valid time. The ocean state is prescribed rather than coupled, varying only linearly between successive monthly anchors without sub-monthly anomalies and remaining identical across all members of a case.

Within each conditioning mode, all 1000 members share identical boundary conditioning and differ only in their latent noise realization. Members are generated one at a time, and member $k$ ($k = 0, \dots, 999$) is seeded with a per-case base seed of zero plus $k$. The base seed is shared across conditioning modes so that member $k$ starts from the same initial latent draw across start-, final-, and start-and-final-conditioned ensembles. All reported evaluations treat each ensemble independently.

\noindent\textbf{Text S2. EOF modes, regression, and cross-validation}

\noindent\textit{Terminology.} Decomposing the free-end $z_{500}$ field yields spatial loading patterns and per-member projection scores. \textbf{EOF$n$} denotes the spatial pattern: one loading value per grid point, displayed in the loading maps. \textbf{PC$n$} denotes the per-member score: the scalar projection of an ensemble member onto EOF$n$. All ensemble-member statistics---correlations with storm position, member percentiles, decile composites, and multiple linear regressions---are evaluated across PC$n$ scores rather than loading maps. When the text describes a mode as ordering or organizing members, it refers to PC$n$; spatial features and extrema refer to EOF$n$. Quoted variance percentages apply to the EOF/PC mode pair.

\noindent\textit{Definition.} Let $N$ be the number of ensemble members and $M$ the number of model pixels inside the analysis domain. For an $N \times M$ member-by-pixel matrix $\mathbf{X}$, ensemble-mean vector $\overline{\mathbf{x}}$, and length-$N$ ones vector $\mathbf{1}$, the singular value decomposition of the member-centered matrix is
\begin{equation}
\mathbf{X} - \mathbf{1}\,\overline{\mathbf{x}}^{\top} = \mathbf{U}\mathbf{S}\mathbf{V}^{\top},
\qquad
\mathrm{EOF}_{n} = \mathbf{V}\mathbf{e}_{n},
\qquad
\mathrm{PC}_{n} = \mathbf{U}\mathbf{S}\mathbf{e}_{n},
\end{equation}
where $\mathbf{e}_{n}$ is the $n$th standard basis vector. Here $\mathrm{EOF}_{n}$ is an $M$-element loading vector across pixels and $\mathrm{PC}_{n}$ is an $N$-element score vector across members. The percentage of variance explained by mode $n$ is
\begin{equation}
f_{n} = s_{n}^{2} \Big/ \sum_{j} s_{j}^{2},
\end{equation}
where $s_{n}$ is the $n$th singular value and the denominator sums over the full spectrum. Loadings are reported in physical units (metres of geopotential height): the decomposition is performed on a field rescaled to unit variance and the uniform scaling factor is restored in displayed maps, altering neither the spatial patterns nor member ordering.

\noindent\textit{EOF construction.} Decompositions are computed across ensemble members over the regional domain specified for each case. Three computational choices follow directly from the data representation: (1) calculations use the model's native equal-area HEALPix pixels, requiring no $\cos(\phi)$ latitude weighting; (2) no climatology is subtracted because the decomposition's own per-pixel centering already removes any constant offset, so subtracting a climatology first would leave the PC scores and EOF patterns unchanged; and (3) each EOF pattern is oriented so its largest-magnitude loading is positive, establishing consistent percentile signs throughout the text.

\noindent\textit{The observed state in PC space.} To evaluate where observations fall along each mode, ERA5's free-end field $\mathbf{x}_{\mathrm{ERA5}}$ is projected onto the member-defined basis by centering on the ensemble mean:
\begin{equation}
\mathrm{PC}_{n}^{\mathrm{ERA5}} = \left( \mathbf{x}_{\mathrm{ERA5}} - \overline{\mathbf{x}} \right)^{\top} \mathrm{EOF}_{n} .
\end{equation}
The percentile $p_{n}$ of the observed state along mode $n$ is the percentage of ensemble members whose score does not exceed it:
\begin{equation}
p_{n} = \frac{100}{N} \, \# \left\{ \, i : \mathrm{PC}_{n}(i) \le \mathrm{PC}_{n}^{\mathrm{ERA5}} \, \right\} ,
\end{equation}
evaluated over all $N = 1000$ members without tracking-based exclusions.

\noindent\textit{The members' ellipsoid.} Let $\mathbf{s}$ denote a state's leading eight PC scores, and $\overline{\mathbf{s}}$ and $\mathbf{C}$ the mean and sample covariance of the $N = 1000$ members' scores. The members' ellipsoid is the set of states whose squared Mahalanobis distance from the ensemble mean, $d^{2} = (\mathbf{s} - \overline{\mathbf{s}})^{\top} \mathbf{C}^{-1} (\mathbf{s} - \overline{\mathbf{s}})$, is no larger than the farthest member's. ERA5's $z_{500}$ state therefore lies within it if at least one member is at least as far from the ensemble mean.

\noindent\textit{Regression.} Two complementary regressions are evaluated from the leading eight modes. The spatial regression maps the height anomaly per unit change in the target quantity. After truncating to the leading eight modes, its slope field is
\begin{equation}
b(\mathbf{r}) = \sum_{n=1}^{8} \frac{\mathrm{cov}\left( \mathrm{PC}_{n}, \, T'_{\mathrm{2m}} \right)}{\mathrm{var}\left( T'_{\mathrm{2m}} \right)} \, \mathrm{EOF}_{n}(\mathbf{r}) ,
\end{equation}
expressed in metres per kelvin for the heatwave case. The second regression predicts the target quantity from the leading eight PC scores by ordinary least squares. Because both regressions depend on the same underlying covariances $\mathrm{cov}(\mathrm{PC}_{n}, \cdot)$, the spatial map and predicted-versus-observed scatter represent dual views of a single statistical fit.

\noindent\textit{Cross-validation.} Each target quantity---a storm property for the hurricane cases or land-mean $T_{\mathrm{2m}}$ for the heatwave---is regressed on the leading eight PC scores using ordinary least squares. Cross-validated $R^2$ values are evaluated using a deterministic 10-fold scheme with the $i$-th member in the regression (in member order, counting from zero) assigned to fold $i \bmod 10$, ensuring reproducibility. Predictions for held-out members are generated from fits on the remaining nine folds, and a single pooled $R^2$ is computed:
\begin{equation}
R^{2} = 1 - \sum_{i=1}^{N} \left( y_{i} - \hat{y}_{i} \right)^{2} \Big/ \sum_{i=1}^{N} \left( y_{i} - \overline{y} \right)^{2}
\end{equation}
across all pooled out-of-fold predictions $\hat{y}_{i}$. Where the main text describes a weak relationship as statistically robust, this refers to the $F$-test on the corresponding in-sample regression with eight predictors, whose $p$-values remain below $10^{-10}$ even when explained variance is under 10\%.

\noindent\textbf{Text S3. Storm tracking and identification}

Storm tracks are extracted using TempestExtremes (Ullrich et al., 2021) applied to the $0.25^{\circ}$-regridded fields. \texttt{DetectNodes} identifies storm candidates as mean sea level pressure (MSLP) minima enclosed by a 200~Pa closed contour within $5.5^{\circ}$ (merging candidates within $6^{\circ}$), requiring a closed warm-core thickness contour with $z_{300}-z_{500}$ decreasing by $\ge 58.8\text{ m}^{2}\text{ s}^{-2}$ within $6.5^{\circ}$ to filter out extratropical lows (Zarzycki \& Ullrich, 2017). \texttt{StitchNodes} connects candidates into trajectories within an $8^{\circ}$ search range, requiring tracks to persist for $\ge 54$ of the 66-hour window with no gap exceeding 24 hours, and to satisfy a 10~m~s$^{-1}$ wind threshold, $\le 50^{\circ}$ latitude bound, and 150~m surface-elevation bound for at least ten timesteps.

Because the global tracker identifies all regional systems, each case defines a focused track domain ($85^{\circ}$--$60^{\circ}$~W, $25^{\circ}$--$45^{\circ}$~N for Superstorm Sandy; $90^{\circ}$--$70^{\circ}$~W, $20^{\circ}$--$36^{\circ}$~N for Hurricane Ian). Each member's storm of interest is resolved by anchoring to the pinned state rather than tracker order: candidate tracks must pass within 500~km of the observed ERA5 storm fix at the pinned frame ($\pm 2$ frames allowed), breaking ties by selecting the path with the most in-domain fixes. A member is deemed analyzable when its resolved trajectory contains at least three valid in-domain fixes and a fix at the free-end frame, uniquely identifying one path per analyzable member.

Ensemble members that are not analyzable are excluded. Exclusion occurs if the tracker does not detect an in-domain track, a track that reaches the pinned frame, three fixes in the domain, or a fix at the free-end frame. However, every ensemble member with a detected in-domain track passes the three-fix threshold and reaches the pinned frame. The number of excluded members is 212 of 1000 for final-conditioned Sandy and 16 of 1000 for start-conditioned Sandy (leaving 788 and 984 analyzable members), and 53 and 243 of 1000 for final- and start-conditioned Ian (leaving 947 and 757 analyzable members). In each case, a large majority of excluded members are excluded only because they lack a free-end fix detected by the tracker. This accounts for 168 out of 212 excluded final-conditioned Sandy members, 15 out of 16 excluded start-conditioned Sandy members, 46 out of 53 excluded final-conditioned Ian members, and 208 out of 243 excluded start-conditioned Ian members.

Free-end storm properties are extracted from the resolved path at exactly the free-end frame. Every storm statistic for Superstorm Sandy and Hurricane Ian in the text (e.g. regressions, central-pressure percentiles, and the landfall split) uses only the analyzable members (i.e. the members whose storm is detected by the tracker at the free-end frame). However, statistics regarding the $z_{500}$ field (EOF variance fractions, ERA5's PC percentiles, and spread) use all 1000 members.

For final-conditioned Ian, the Florida-versus-bypass bifurcation evaluates the resolved path against a geographic bounding box: members with any fix inside $83.5^{\circ}$--$80.0^{\circ}$~W, $24.5^{\circ}$--$30.0^{\circ}$~N (spanning southwest Florida landfall near Cayo Costa across the peninsula) are classified as Florida landfalls, and bypass members otherwise. Because all analyzable members are anchored to the final South Carolina landfall, bypass tracks represent counterfactual trajectories reaching Ian's observed state without traversing Florida. The South Carolina box in the main text ($81.5^{\circ}$--$78.5^{\circ}$~W, $32.0^{\circ}$--$34.5^{\circ}$~N) highlights the terminal anchoring region and does not participate in classification.

\noindent\textbf{Text S4. Free-end $z_{500}$ ensemble spread}

To compare ensemble dispersion across conditioning directions, the free-end $z_{500}$ spread is evaluated over the synoptic analysis domain. For member $i$ with free-end height $z_{500}^{(i)}(\mathbf{r}_{m})$ and ensemble mean $\overline{z}_{500}(\mathbf{r}_{m})$ at the location $\mathbf{r}_{m}$ of pixel $m$, the domain-mean standard deviation is
\begin{equation}
\sigma = \left[ \frac{1}{M} \sum_{m=1}^{M} \frac{1}{N-1} \sum_{i=1}^{N} \left( z_{500}^{(i)}(\mathbf{r}_{m}) - \overline{z}_{500}(\mathbf{r}_{m}) \right)^{2} \right]^{1/2} ,
\end{equation}
computed by averaging sample variance per pixel across all $M$ domain pixels before taking the square root. All $N = 1000$ members enter the calculation over native equal-area HEALPix pixels, requiring no latitude weighting.

The main text reports the free-end spread ratio between final- and start-conditioned ensembles $\sigma_{\mathrm{final}} / \sigma_{\mathrm{start}}$: 0.84 for the 2021 PNW heatwave (15.7 versus 18.8~m), 0.86 for Hurricane Ian (12.3 versus 14.3~m), and 0.89 for Superstorm Sandy (23.5 versus 26.4~m). Because the free end represents the unconditioned boundary of each 66-hour window, the compared spreads sit at identical lead times from their pinned frames but at valid times 66 hours apart.

\noindent\textbf{Text S5. The 500~hPa ageostrophic fraction}

Figures S15 to S17 evaluate how much of the generated 500~hPa flow deviates from geostrophic balance. This diagnostic is computed directly on the generated fields without requiring observed outcomes; ERA5 serves as an observational reference.

Geostrophic wind components are computed from the 500~hPa geopotential $\Phi$ (in m$^{2}$~s$^{-2}$):
\begin{equation}
u_{g} = -\frac{1}{fR}\frac{\partial \Phi}{\partial \phi},
\qquad
v_{g} = \frac{1}{fR\cos\phi}\frac{\partial \Phi}{\partial \lambda},
\qquad
f = 2\Omega\sin\phi ,
\end{equation}

where $\phi$ is latitude, $\lambda$ is longitude, $R$ is Earth's radius, $\Omega$ is Earth's rotation rate, $f = 2\Omega\sin\phi$ is the Coriolis parameter, and spatial derivatives are evaluated via centered finite differences on the $0.25^{\circ}$ latitude--longitude grid (periodic in longitude). The ageostrophic wind is defined as the residual $\mathbf{V}_{a} = \mathbf{V} - \mathbf{V}_{g}$ between the horizontal wind $\mathbf{V}$ and the geostrophic wind $\mathbf{V}_{g} = (u_{g}, v_{g})$, capturing all flow components not accounted for by geostrophic balance, including flow curvature.

Wind speeds are averaged over the $30^{\circ}$--$60^{\circ}$~$|\phi|$ band in both hemispheres across all longitudes (242 latitude rows by 1440 longitudes, totalling 348,480 points per frame) using $\cos\phi$ area weighting, accounting for the 1.7 factor decrease in grid-cell area across the band. Denoting the area-weighted domain mean by $\langle \cdot \rangle$, the ageostrophic fraction is
\begin{equation}
f_{a} = \frac{\langle |\mathbf{V}_{a}| \rangle}{\langle |\mathbf{V}| \rangle} ,
\end{equation}
evaluated as a ratio of domain means to avoid domination by light-wind cells. Within this midlatitude band, Coriolis parameter $f$ remains far from zero. Because these scalar speeds represent domain averages of vector magnitudes, $\langle |\mathbf{V}_{g}| \rangle$ and $\langle |\mathbf{V}_{a}| \rangle$ do not sum linearly to $\langle |\mathbf{V}| \rangle$.

A fixed midlatitude band is used rather than a storm-centric box because the latter would fail to support consistent cross-member averaging when storm locations diverge across members; the fixed band ensures uniform spatial sampling across all members.

Three reference curves are shown in each panel. Because model fields are generated on native HEALPix pixels and regridded to latitude--longitude, whereas ERA5 is native latitude--longitude, the two differ in interpolation history. The dotted green line represents ERA5 subjected to the same latitude--longitude~$\rightarrow$~HEALPix~$\rightarrow$~latitude--longitude round-trip interpolation applied to conditioning frames. At pinned frames, the green curve provides the exact counterpart to the input data. At unpinned frames, it provides an upper bound on regridding-induced smoothing rather than an exact counterpart, because generated fields undergo one interpolation pass whereas the round trip undergoes two; the correctly smoothed reference for those frames lies between the two ERA5 curves. A reader comparing the ensemble against the dashed black line alone is comparing across that difference in interpolation history, which the green curve exists to bracket.

\clearpage

%% ------------------------------------------------------------------------ %%
%  FIGURES
%% ------------------------------------------------------------------------ %%

\begin{figure}
\centering
% 0.82\textwidth, not \textwidth: this figure is tall (h/w = 0.78) and AGU's
% draft mode is double-spaced, so at full width the float overruns the page
% and the caption collides with the footer.
\includegraphics[width=0.82\textwidth]{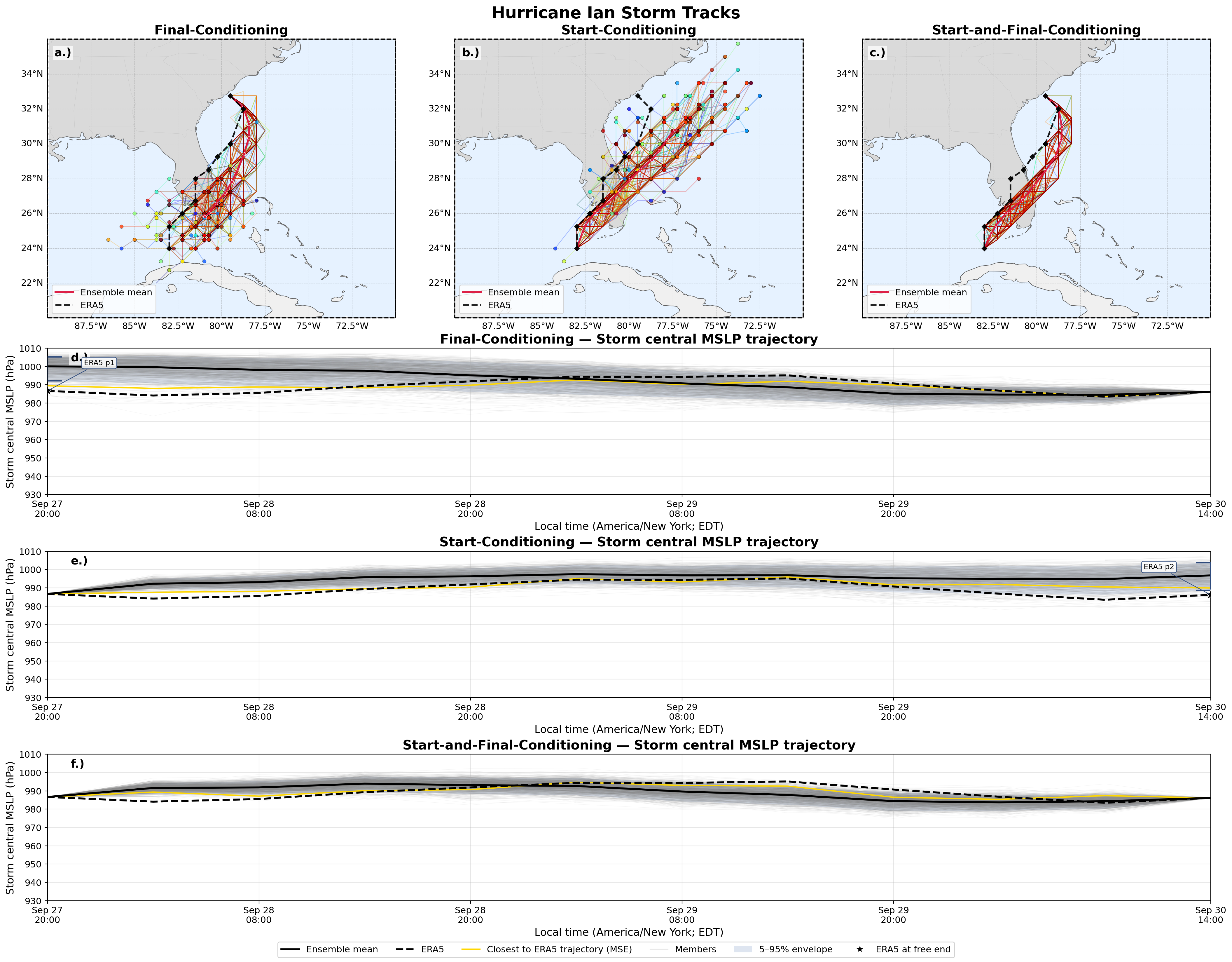}
\caption{Hurricane Ian tracks and central-pressure trajectories for 1000-member ensembles under final- (a,d), start- (b,e), and start-and-final- (c,f) conditioning. (a--c) Storm-center tracks connect 6-hourly TempestExtremes fixes, effectively resolved at the native HPX64 ($\sim$100~km) scale; the $0.25^{\circ}$ regridding used for tracking adds no finer-scale information. The angularity is a discrete-sampling artifact, most pronounced where the pressure minimum is weak or diffuse, and does not reflect physical track variability. (d--f) Resolved storm central pressure for individual members (gray), ensemble mean (solid black), ERA5 (dashed black), and minimum-MSE member (gold). Blue whiskers at the free-end margin in (d,e) denote the 5th--95th percentile envelope, annotated with ERA5's percentile among analyzable members: 1.0 at the final-conditioned free end and 1.8 at the start-conditioned free end. Start-and-final-conditioning (c,f) pins both boundaries and carries no free end.}
\label{fig:ian_tc_tracks}
\end{figure}

% Figures S2-S5 carry the four hurricane track and intensity regressions.
% All four share a layout, so they share the 0.82\textwidth sizing note above:
% h/w is 0.79-0.80 here against ian_tc_tracks.png's 0.78.
\begin{figure}
\centering
\includegraphics[width=0.82\textwidth]{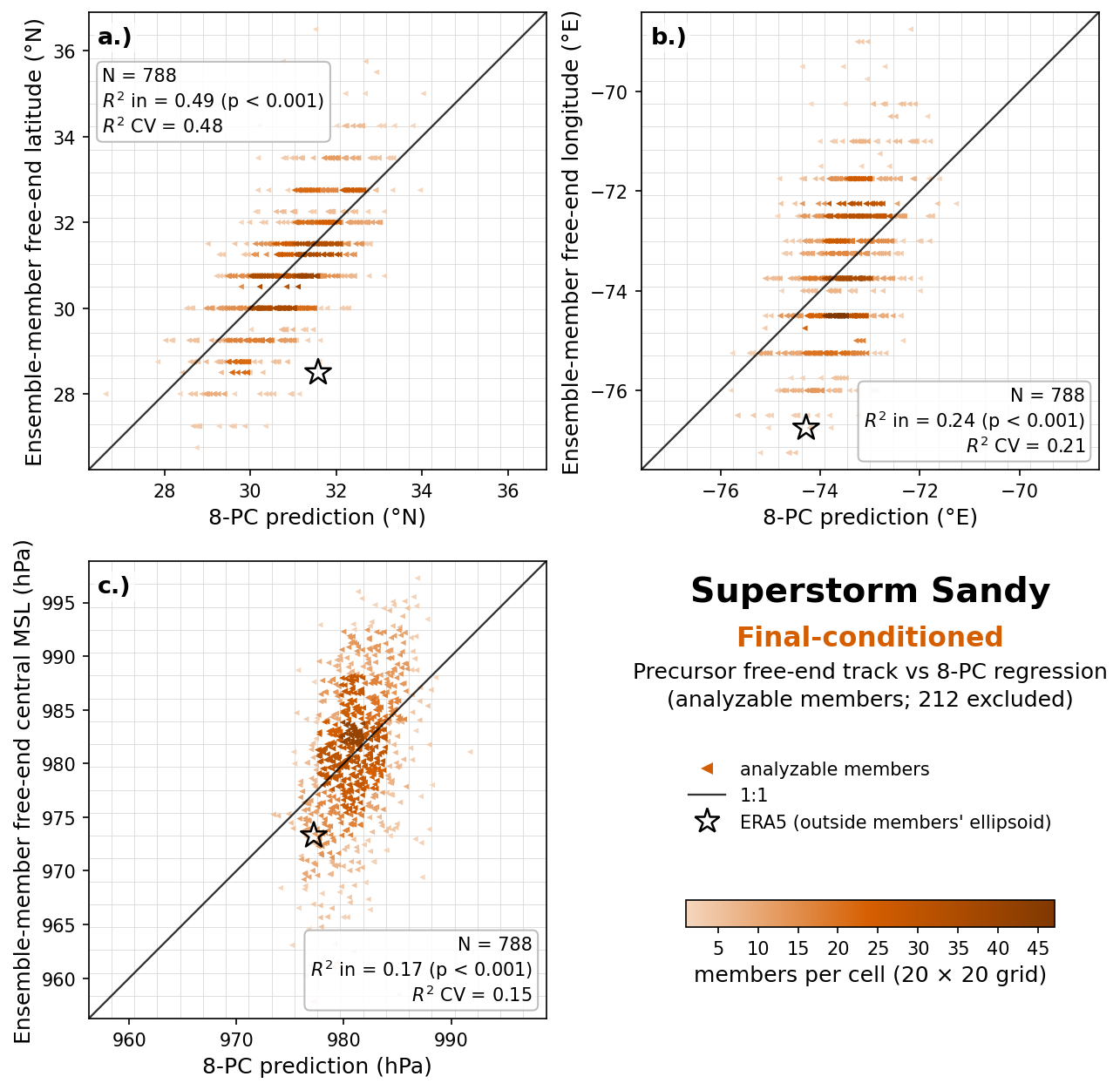}
\caption{Predicted versus ensemble-member free-end storm properties for the final-conditioned Superstorm Sandy ensemble ($N=788$ analyzable members; 212 excluded) at the precursor boundary (October 27, 2012, 06:00 UTC): (a) latitude, (b) longitude, and (c) central MSLP. Predictions are from ordinary least-squares regressions on the leading eight principal components of the free-end $z_{500}$ field ($100^{\circ}$--$30^{\circ}$~W, $25^{\circ}$--$75^{\circ}$~N); insets report in-sample and 10-fold cross-validated $R^2$. The star marks ERA5 evaluated at its observed $z_{500}$ state. This star is filled if that state lies within the members' ellipsoid and hollow otherwise (Text~S2). Members are vermilion left-pointing triangles (final-conditioned) or blue right-pointing triangles (start-conditioned) and are shaded by the number of members in each cell of the faint $20 \times 20$ grid. The rows in (a) and (b) reflect the ${\sim}100$~km native grid on which storm positions are resolved (Figure~S1).}
\label{fig:sandy_track_regression_end}
\end{figure}

\begin{figure}
\centering
\includegraphics[width=0.82\textwidth]{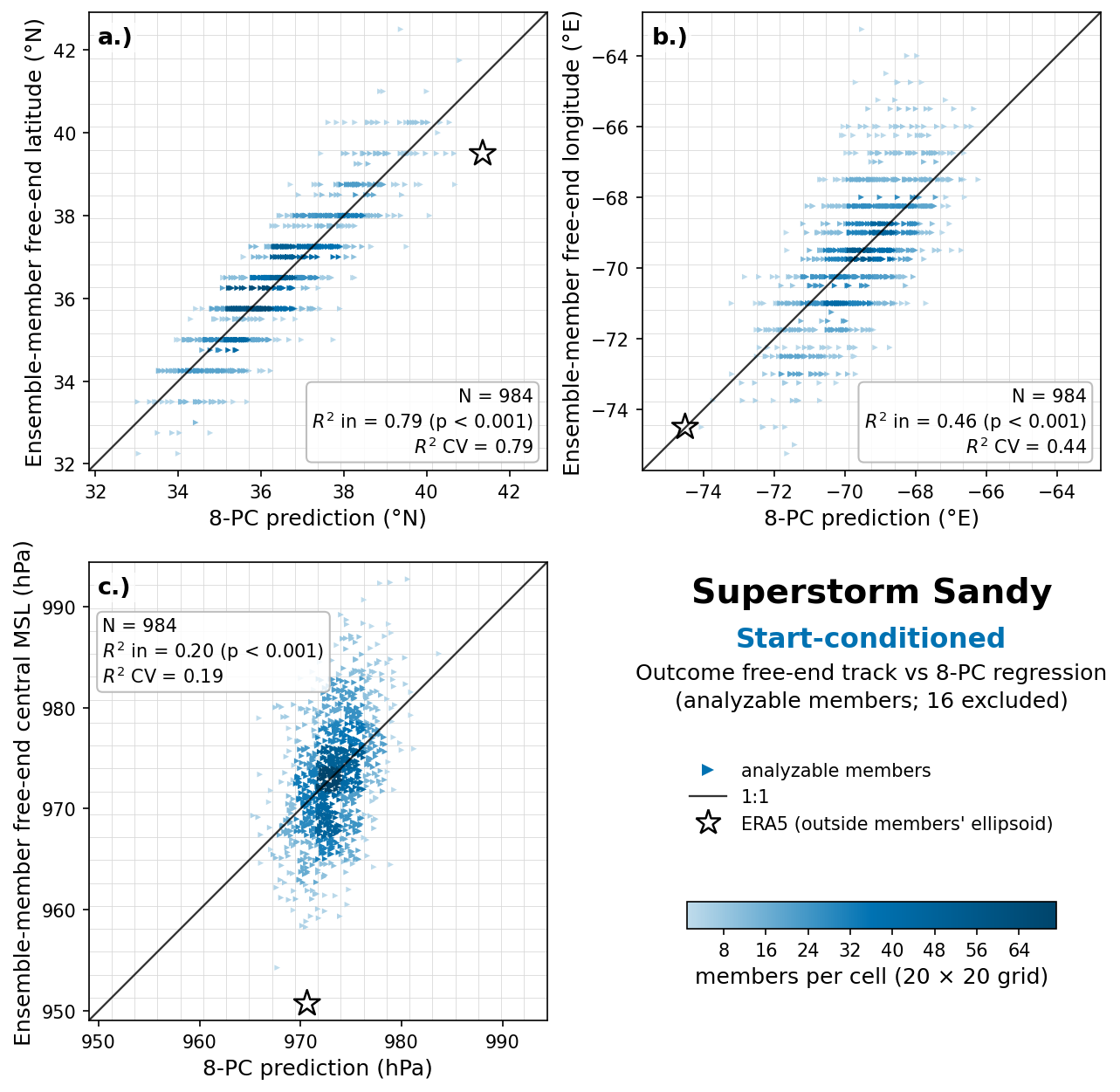}
\caption{Predicted versus ensemble-member free-end storm properties for the start-conditioned Superstorm Sandy ensemble ($N=984$ analyzable members; 16 excluded) at the outcome boundary (October 30, 2012, 00:00 UTC), analogous to Figure~S2. Predictions are from ordinary least-squares regressions on the leading eight principal components of the free-end $z_{500}$ field ($100^{\circ}$--$30^{\circ}$~W, $25^{\circ}$--$75^{\circ}$~N).}
\label{fig:sandy_track_regression_start}
\end{figure}

\begin{figure}
\centering
\includegraphics[width=0.82\textwidth]{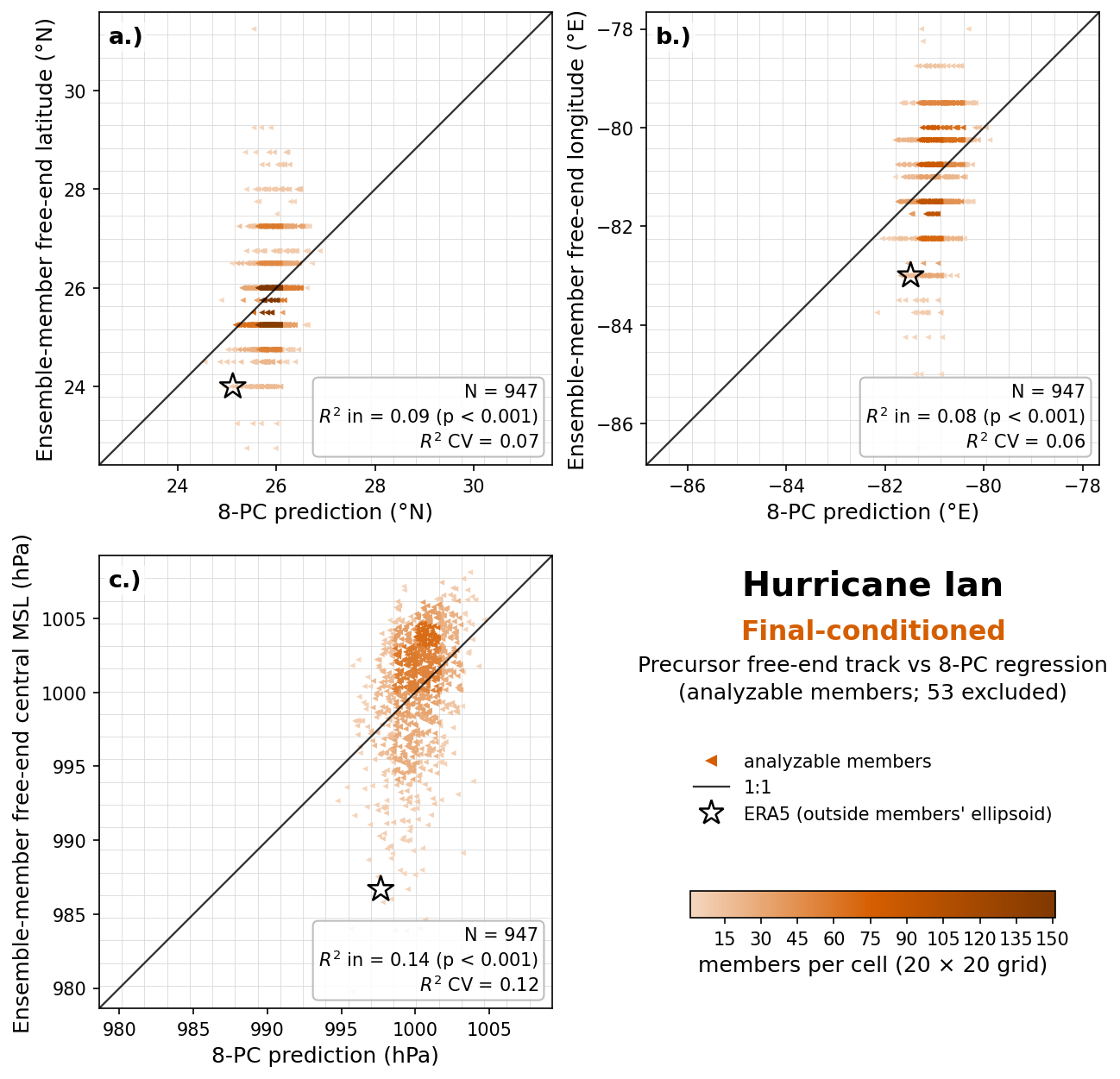}
\caption{Predicted versus ensemble-member free-end storm properties for the final-conditioned Hurricane Ian ensemble ($N=947$ analyzable members; 53 excluded) at the precursor boundary (September 28, 2022, 00:00 UTC), analogous to Figure~S2. Predictions are from ordinary least-squares regressions on the leading eight principal components of the free-end $z_{500}$ field ($110^{\circ}$--$50^{\circ}$~W, $15^{\circ}$--$50^{\circ}$~N).}
\label{fig:ian_track_regression_end}
\end{figure}

\begin{figure}
\centering
\includegraphics[width=0.82\textwidth]{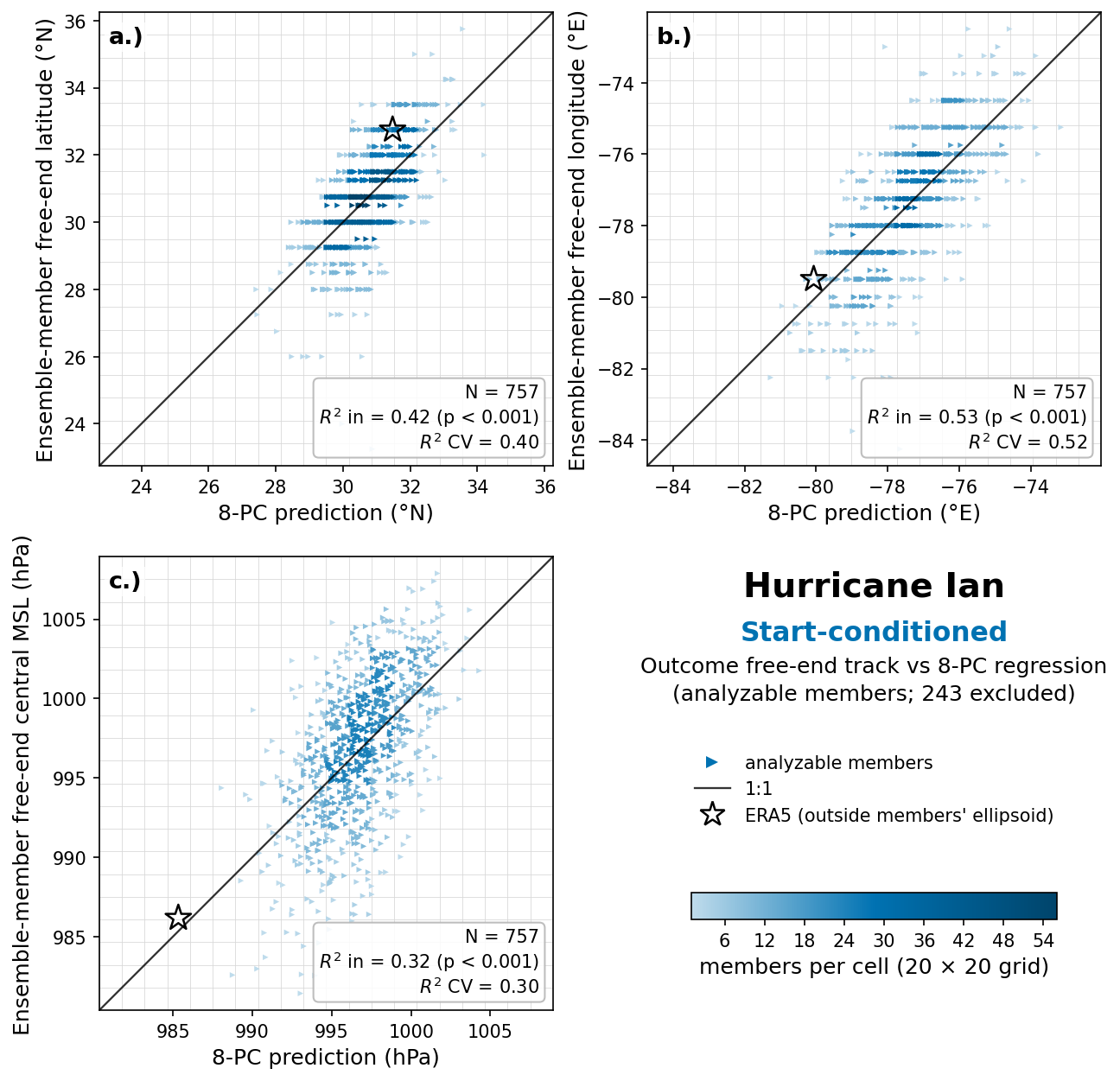}
\caption{Predicted versus ensemble-member free-end storm properties for the start-conditioned Hurricane Ian ensemble ($N=757$ analyzable members; 243 excluded) at the outcome boundary (September 30, 2022, 18:00 UTC), analogous to Figure~S2. Predictions are from ordinary least-squares regressions on the leading eight principal components of the free-end $z_{500}$ field ($110^{\circ}$--$50^{\circ}$~W, $15^{\circ}$--$50^{\circ}$~N).}
\label{fig:ian_track_regression_start}
\end{figure}

% Figures S6-S11 are the spread/RMSE/CRPS verification battery, one per
% (case, variable) pair. Each PNG is already a complete five-panel figure, so
% they are not composed further: merging a case's two variables into one float
% would duplicate the (a)-(e) panel letters. These are taller than the
% regression figures (h/w = 0.90 against 0.79-0.80), so 0.72\textwidth here
% gives the same rendered height that 0.82\textwidth gives them.
\begin{figure}
\centering
\includegraphics[width=0.72\textwidth]{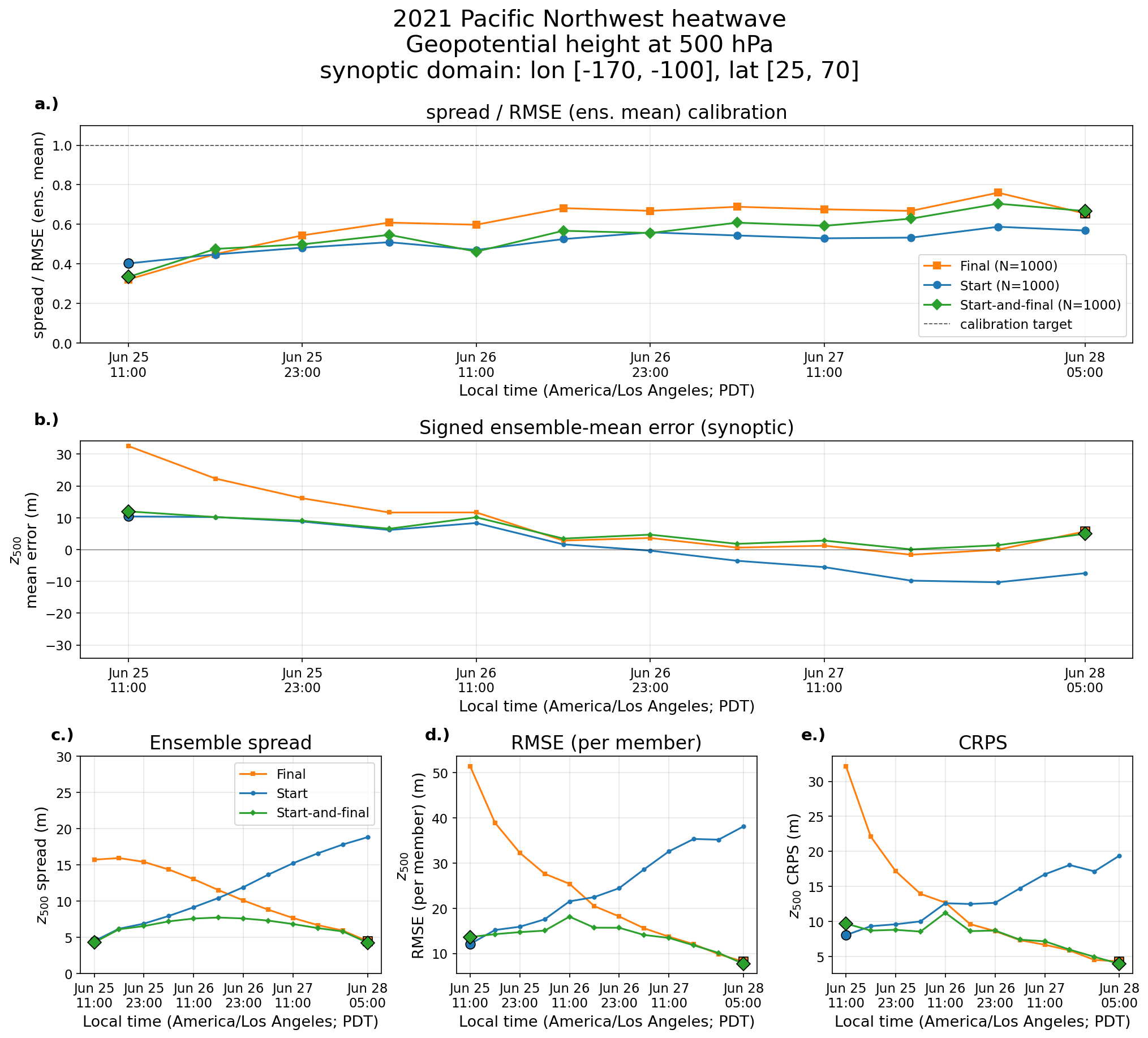}
\caption{Ensemble spread, ensemble-mean error, and probabilistic skill against ERA5 as a function of lead time for the 2021 Pacific Northwest heatwave $z_{500}$ field over the synoptic domain ($170^{\circ}$--$100^{\circ}$~W, $25^{\circ}$--$70^{\circ}$~N), across all three conditioning modes ($N = 1000$ each). (a) Ratio of ensemble spread to ensemble-mean RMSE (dashed horizontal line at unity denotes statistical consistency; values below unity indicate under-dispersion). (b) Signed ensemble-mean error. (c) Domain-mean ensemble spread (Equation~S7). (d) Per-member RMSE. (e) Continuous ranked probability score (CRPS). Enlarged markers denote pinned boundaries (start, final, or start-and-final). Quantities are domain-averaged over native equal-area HEALPix pixels without latitude weighting. Free-end spreads in (c)---15.7~m final-conditioned and 18.8~m start-conditioned---form the 0.84 ratio reported in Text~S4.}
\label{fig:spread_rmse_crps_pnw_heatwave_z500}
\end{figure}

\begin{figure}
\centering
\includegraphics[width=0.72\textwidth]{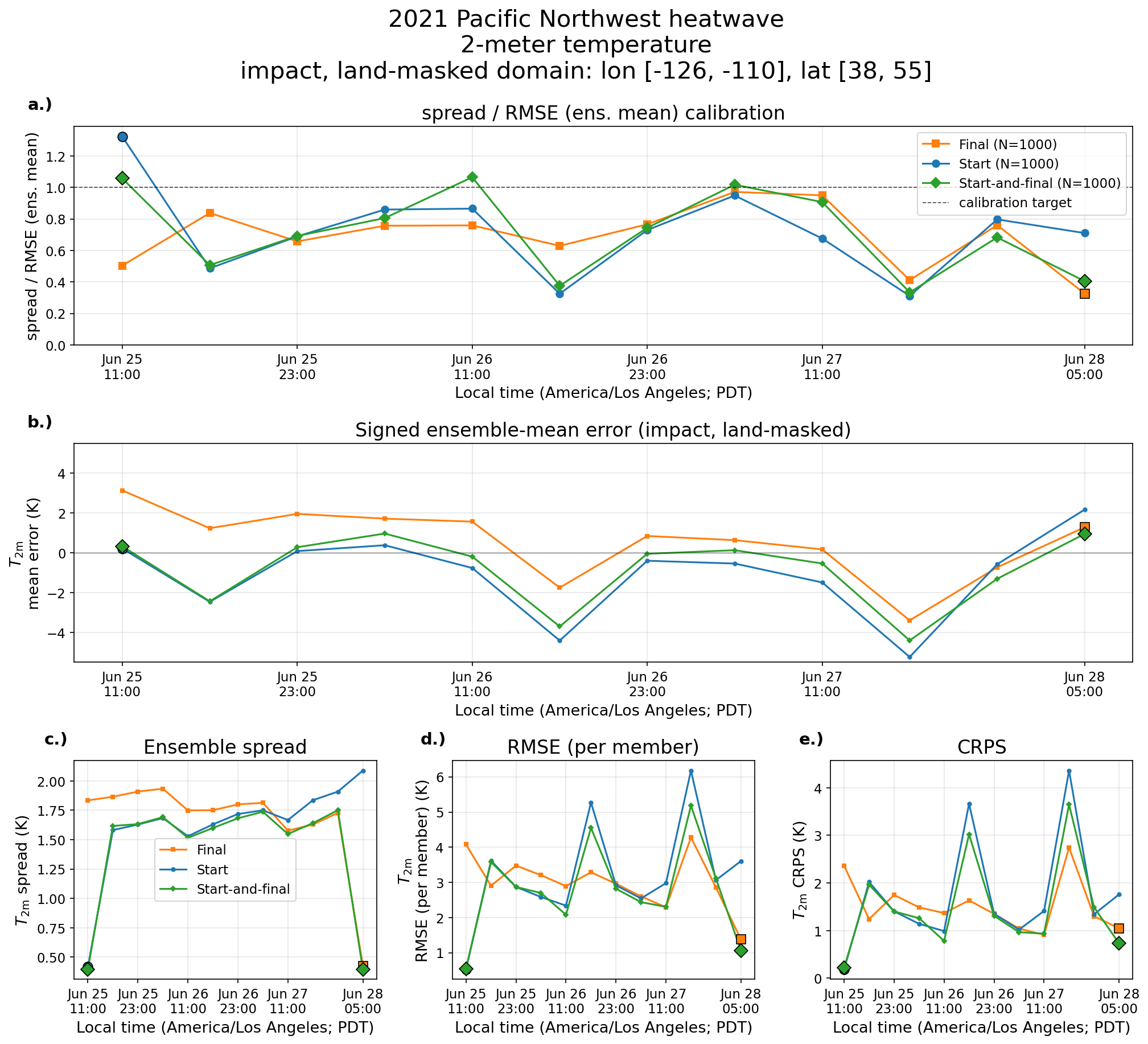}
\caption{As in Figure~S6, but for 2-meter temperature over land pixels of the impact domain ($126^{\circ}$--$110^{\circ}$~W, $38^{\circ}$--$55^{\circ}$~N), using the land mask defined in Section~3.1. Ocean pixels are excluded because they track prescribed, member-identical sea surface temperatures. Unlike the $z_{500}$ and MSLP fields of Figures~S6 and S8--S11, the spread-to-error ratio in (a) reaches and exceeds unity across several frames, demonstrating that under-dispersion is not universal across all variables.}
\label{fig:spread_rmse_crps_pnw_heatwave_t2m}
\end{figure}

\begin{figure}
\centering
\includegraphics[width=0.72\textwidth]{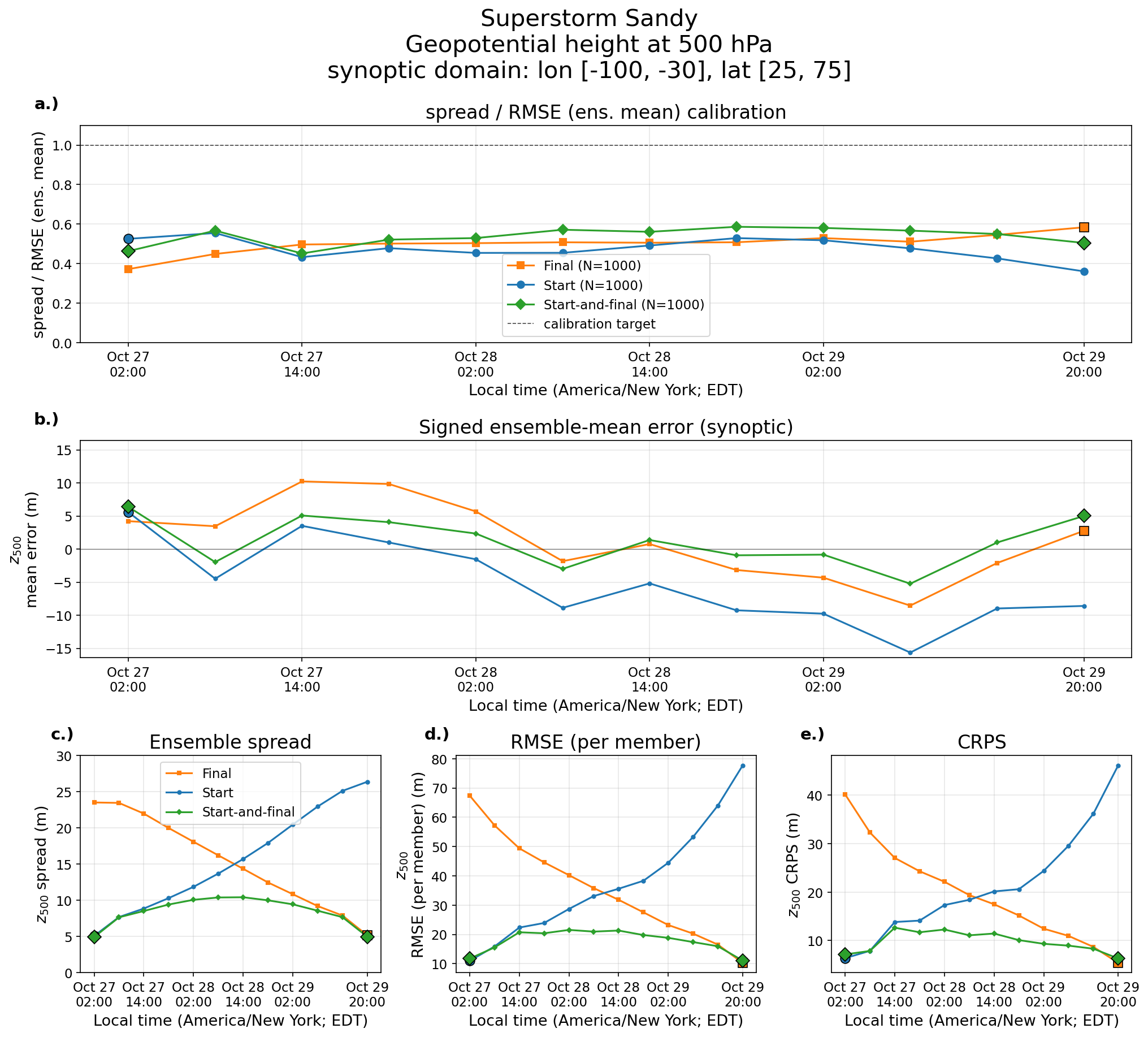}
\caption{As in Figure~S6, but for Superstorm Sandy $z_{500}$ over its synoptic domain ($100^{\circ}$--$30^{\circ}$~W, $25^{\circ}$--$75^{\circ}$~N), matching the domain used for EOF analysis and Text~S4. Free-end spreads in (c)---23.5~m final-conditioned and 26.4~m start-conditioned---form the 0.89 ratio reported in Text~S4.}
\label{fig:spread_rmse_crps_sandy_z500}
\end{figure}

\begin{figure}
\centering
\includegraphics[width=0.72\textwidth]{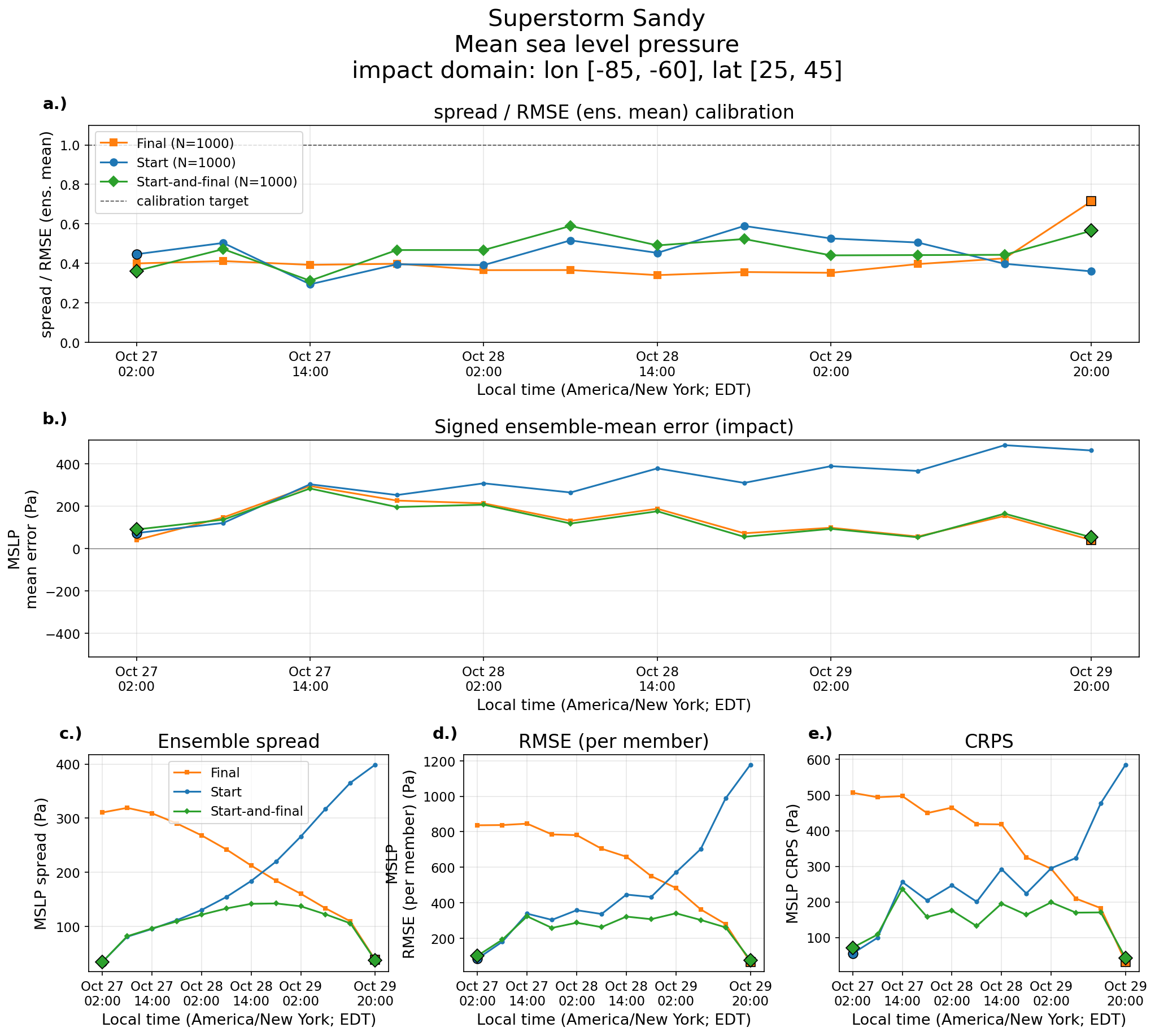}
\caption{As in Figure~S8, but for mean sea level pressure (MSLP) over the cyclone impact domain ($85^{\circ}$--$60^{\circ}$~W, $25^{\circ}$--$45^{\circ}$~N). The free-end spread ratio between final- and start-conditioned ensembles in (c) is 0.78 for MSLP over this localized domain, distinct from the synoptic $z_{500}$ spread ratios defined in Text~S4.}
\label{fig:spread_rmse_crps_sandy_msl}
\end{figure}

\begin{figure}
\centering
\includegraphics[width=0.72\textwidth]{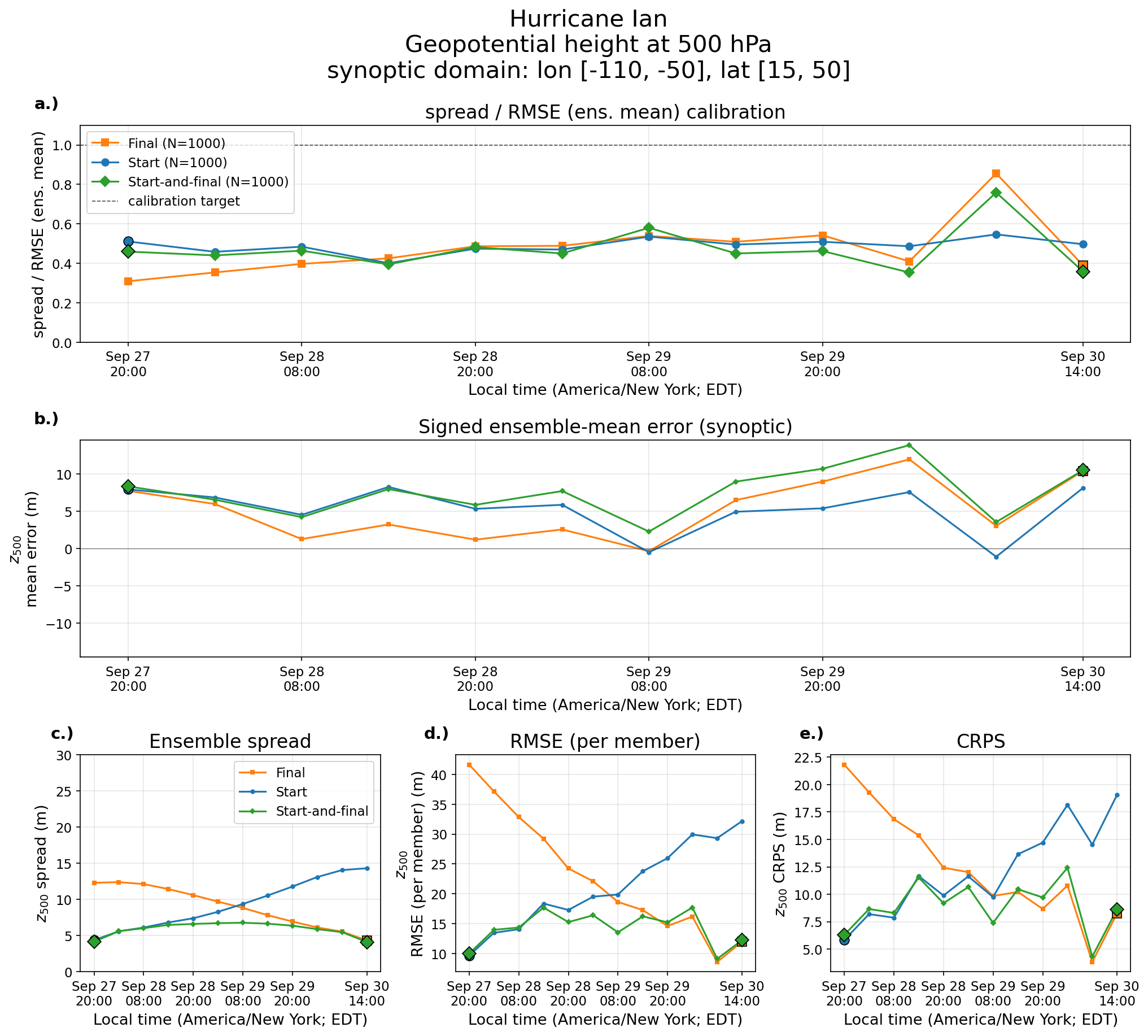}
\caption{As in Figure~S6, but for Hurricane Ian $z_{500}$ over its synoptic domain ($110^{\circ}$--$50^{\circ}$~W, $15^{\circ}$--$50^{\circ}$~N). Free-end spreads in (c)---12.3~m final-conditioned and 14.3~m start-conditioned---form the 0.86 ratio reported in Text~S4.}
\label{fig:spread_rmse_crps_ian_z500}
\end{figure}

\begin{figure}
\centering
\includegraphics[width=0.72\textwidth]{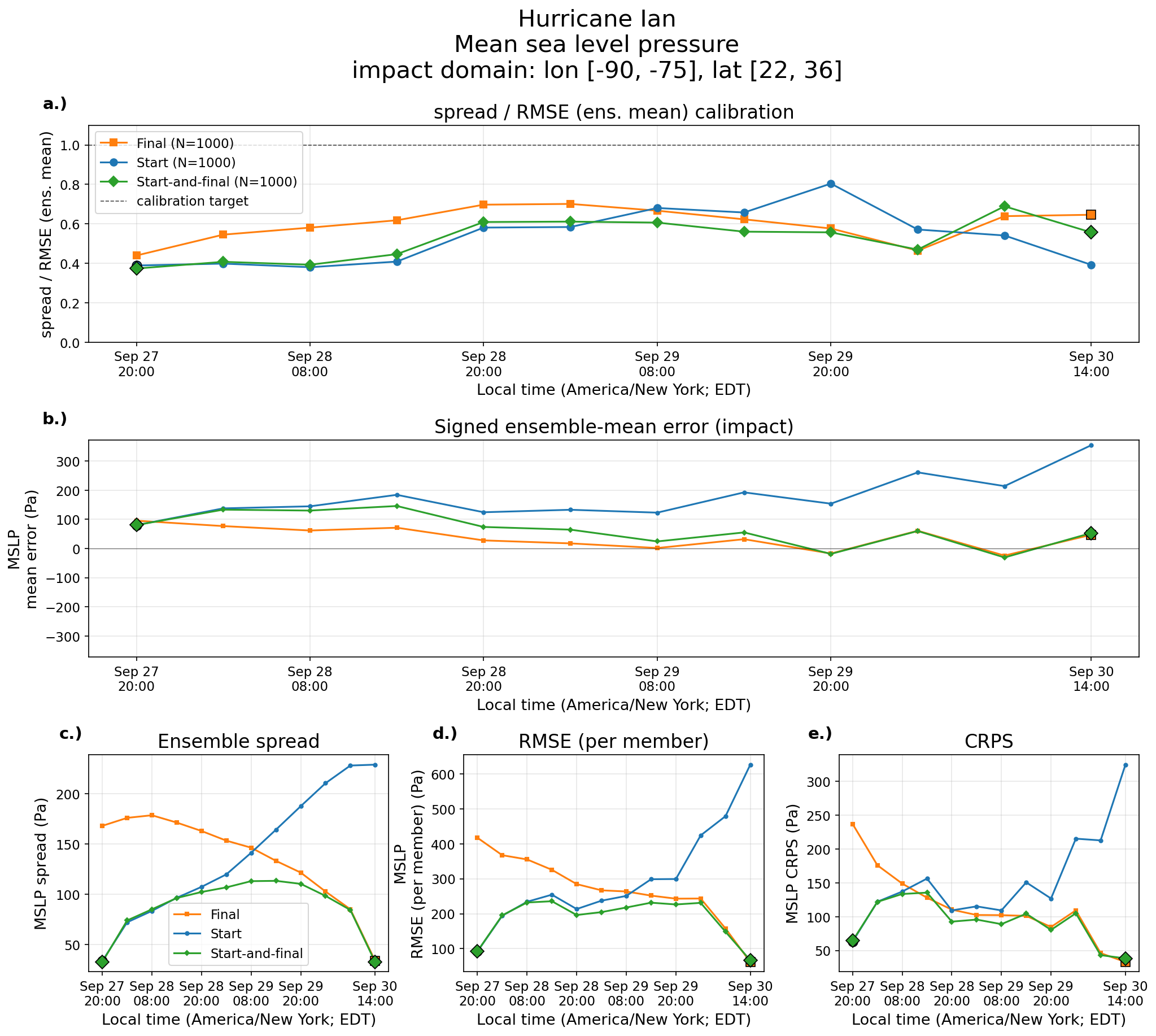}
\caption{As in Figure~S10, but for mean sea level pressure (MSLP) over the cyclone impact domain ($90^{\circ}$--$75^{\circ}$~W, $22^{\circ}$--$36^{\circ}$~N). The free-end spread ratio between final- and start-conditioned ensembles in (c) is 0.75 for MSLP over this impact domain, distinct from the synoptic $z_{500}$ ratio of Text~S4.}
\label{fig:spread_rmse_crps_ian_msl}
\end{figure}

% Figure S12 is the conditioning stress test cited from D3 of the Discussion.
% It is an EXISTENCE demonstration for the admissibility point, not a
% characterization of a boundary, so no member count or random-draw disclaimer
% belongs in the caption (PAPER_PLAN.md, parked-design amendment 2026-08-14).
% Full width because the panels are wide (h/w = 0.64 against the regression
% figures' 0.79-0.80). The caption must disclose the contrivances -- cross-period
% pair, SST clocked to start_time, frame-11 slot/field clock mismatch -- because
% the panel headers make the year jump between F10 and F11 visible to any reader.
\begin{figure}
\centering
\includegraphics[width=\textwidth]{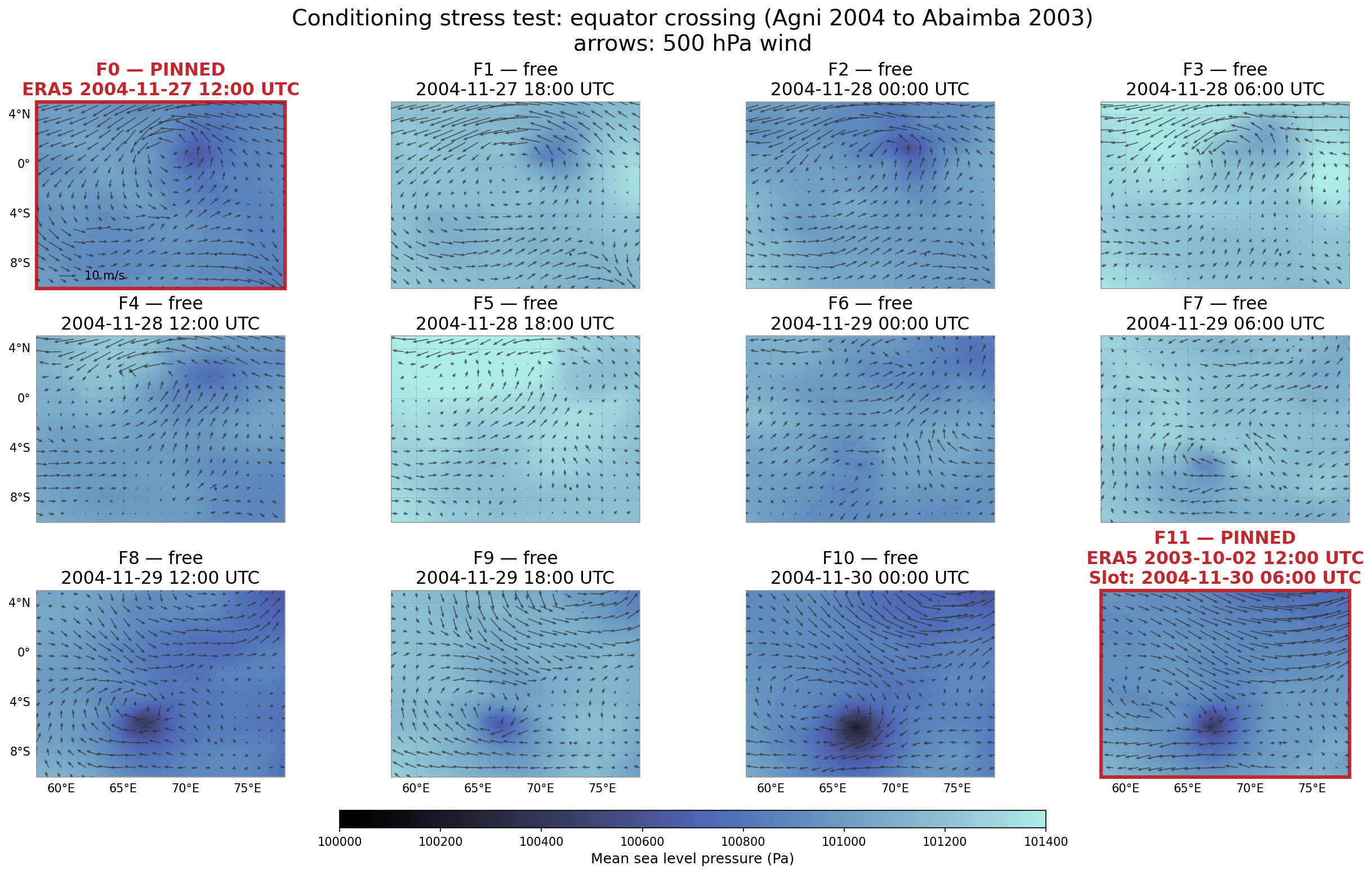}
\caption{This figure demonstrates a start-and-final-conditioning stress test in which the start and final conditioning frames (marked with red outlines) show tropical cyclones that are positioned on opposite sides of the equator. Since cBottle-video does not reproduce conditioned frames exactly, all 12 frames are generated output. Frame 0 shows Cyclone Agni (located at $0.5^{\circ}$~N, $70.5^{\circ}$~E) and is pinned to the state of the global atmosphere on 27 November 2004, 12:00 UTC. Frame 11 shows Tropical Storm Abaimba (located at $5.8^{\circ}$~S, $67.0^{\circ}$~E) and is pinned to the state of the global atmosphere on 2 October 2003, 12:00 UTC. The calendar embedding aligns with a 66-hour sequence beginning at the timestamp used for the first frame (i.e. 27 November 2004, 12:00 UTC). This experiment evaluates conditioning admissibility rather than forecast skill.}
\label{fig:conditioning_stress_test_equator}
\end{figure}

\begin{figure}
\centering
\includegraphics[width=\textwidth]{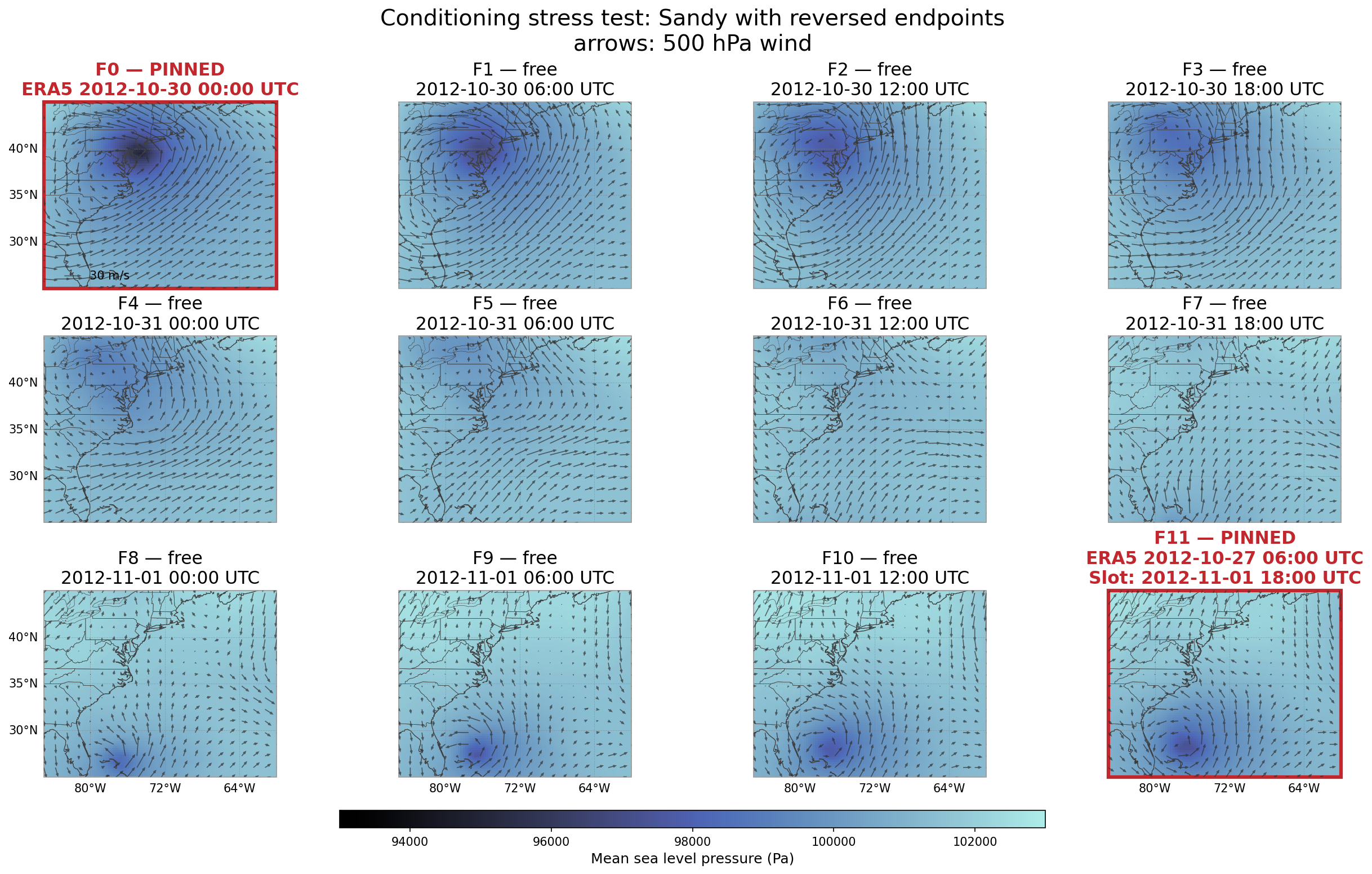}
\caption{This figure demonstrates a start-and-final-conditioning stress test in which the start and final conditioning frames (marked with red outlines) are the swapped endpoints used for start-and-final-conditioning in Figure 3c. In other words, the 66-hour trajectory begins at Sandy's landfall (30 October 2012, 00:00 UTC) and ends at Sandy's state 66 hours before landfall (27 October 2012, 06:00 UTC). The calendar embedding aligns with a 66-hour sequence beginning at the timestamp used for the first frame (i.e. Sandy's landfall). Since cBottle-video does not reproduce conditioned frames exactly, all 12 frames are generated output. This experiment evaluates conditioning admissibility rather than forecast skill.}
\label{fig:conditioning_stress_test_reverse_sandy}
\end{figure}

\begin{figure}
\centering
\includegraphics[width=\textwidth]{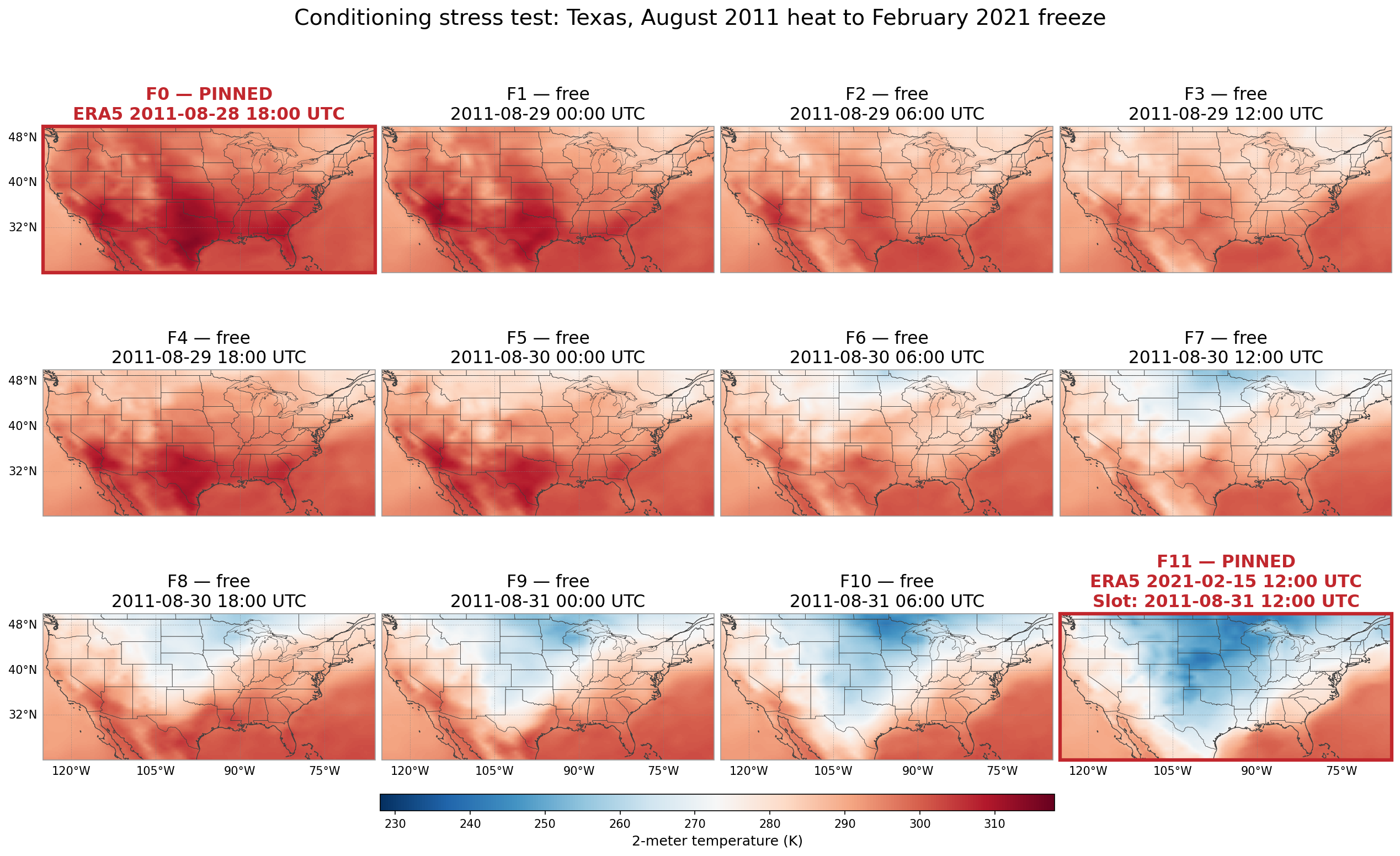}
\caption{This figure demonstrates a start-and-final-conditioning stress test in which the start and final conditioning frames (marked with red outlines) show the contiguous United States on 28 August 2011, 18:00 UTC and 15 February 2021, 12:00 UTC, respectively. This is intended to see if the model can interpolate between an extreme summer heatwave and an extreme winter storm in a 66-hour window. The calendar embedding aligns with a 66-hour sequence beginning at the timestamp used for the first frame (i.e. 28 August 2011, 18:00 UTC). Since cBottle-video does not reproduce conditioned frames exactly, all 12 frames are generated output. This experiment evaluates conditioning admissibility rather than forecast skill.}
\label{fig:conditioning_stress_test_texas}
\end{figure}

% Figures S15-S17 are the ageostrophic-fraction diagnostic, one float per case,
% cited from D4. Each PNG is already a complete twelve-panel figure, so as with
% S6-S11 they are not composed further: merging cases into one float would
% duplicate the (a)-(l) panel letters. h/w = 0.78, matching ian_tc_tracks.png,
% so they take its 0.82\textwidth sizing.
%
% The axis ranges of all three are pinned in scripts/_shared/axis_limits.yaml
% (analysis `ageostrophic`, variable `uvz500`) rather than autoscaled, which is
% what licenses the caption's claim that the three figures may be compared
% directly. If any of these PNGs is ever re-rendered, that block must still be
% in force or the claim becomes false -- the aggregator warns at plot time when
% data exceeds a configured limit, so a silent change is not possible.
\begin{figure}
\centering
\includegraphics[width=0.82\textwidth]{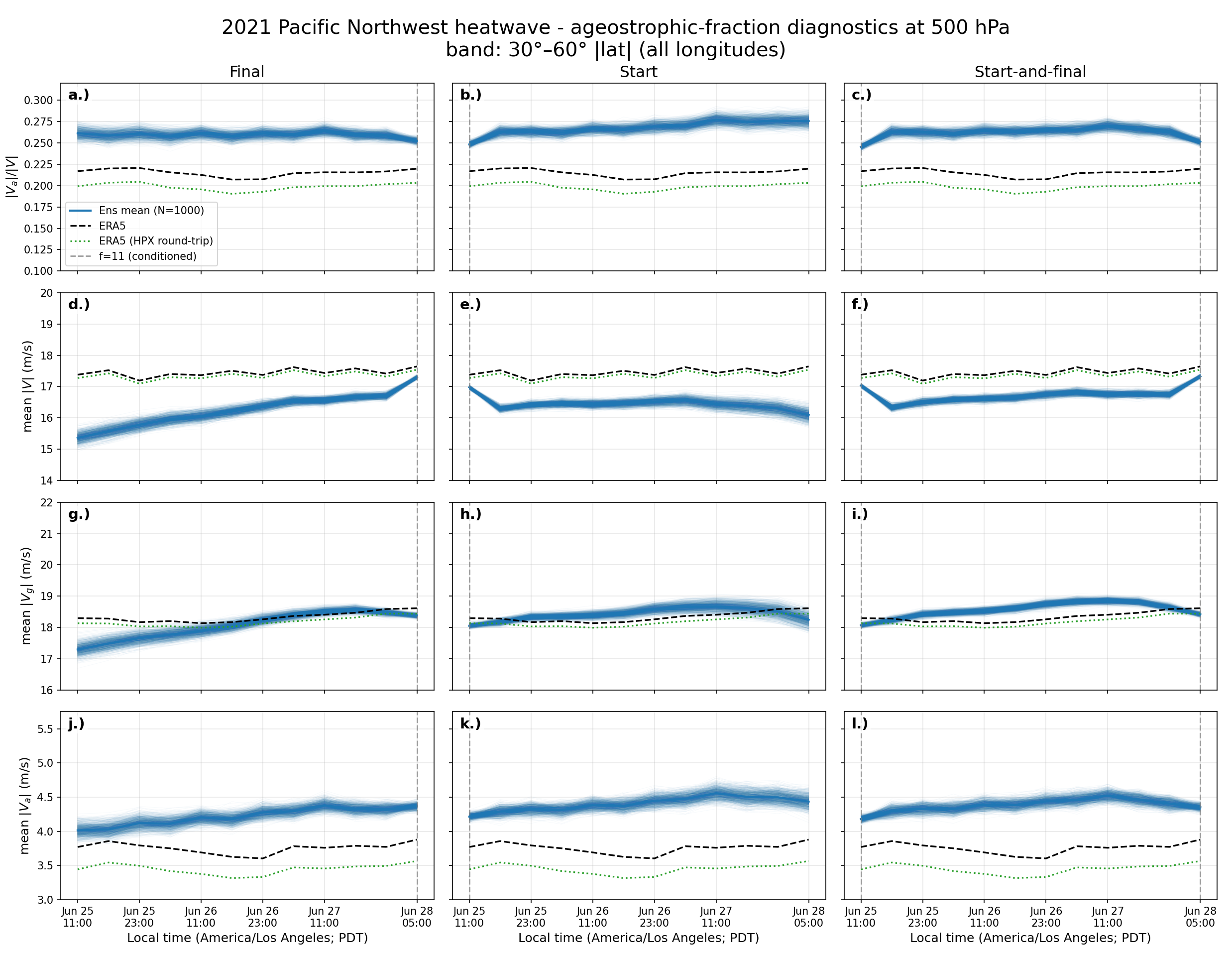}
\caption{Ageostrophic diagnostics at 500~hPa for the 2021 Pacific Northwest heatwave ensembles across all three conditioning modes ($N = 1000$ each), as defined in Text~S5. Rows display the ageostrophic fraction $\langle |\mathbf{V}_{a}| \rangle / \langle |\mathbf{V}| \rangle$ (a--c) and $\cos\phi$-weighted mean speeds $\langle |\mathbf{V}| \rangle$ (d--f), $\langle |\mathbf{V}_{g}| \rangle$ (g--i), and $\langle |\mathbf{V}_{a}| \rangle$ (j--l); columns show final-, start-, and start-and-final-conditioning. Individual ensemble members are shown in thin blue, ensemble means in bold blue, ERA5 in dashed black, and ERA5 passed through the HEALPix round-trip interpolation in dotted green. The green line, not the black, is the reference matching the ensemble's interpolation history (Text~S5); it lies below the black line, so regridding reduces the ageostrophic fraction rather than raising it. Gray dashed vertical lines mark each mode's pinned frames. Common axis ranges allow direct visual comparison across Figures~S15--S17. Mean speeds represent averages of vector magnitudes and are not additive across rows. What is measured is the ageostrophic fraction at one level, over one band, against one balance relation; it is not a general statement about balance in the generated fields.}
\label{fig:ageostrophic_pnw_heatwave}
\end{figure}

\begin{figure}
\centering
\includegraphics[width=0.82\textwidth]{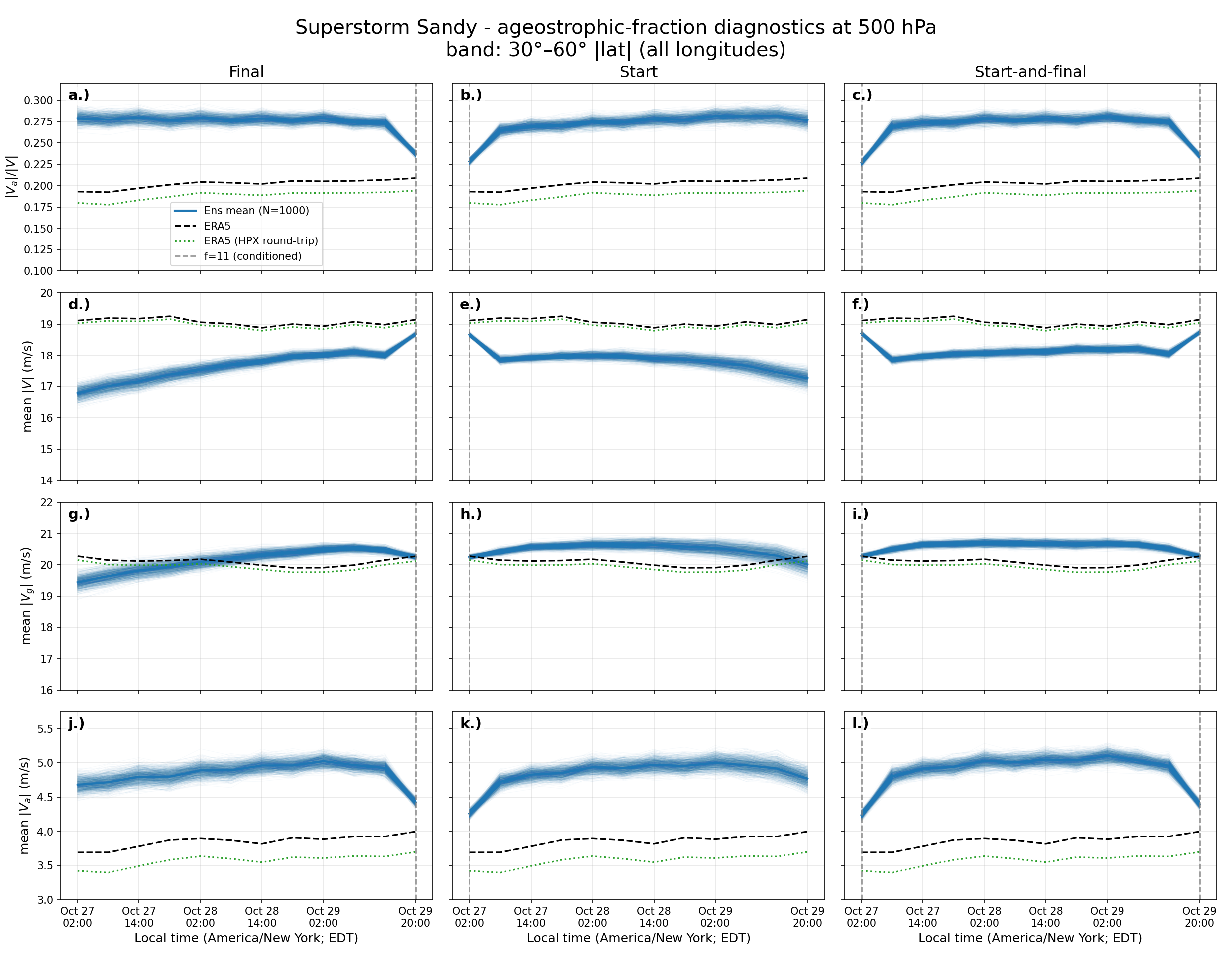}
\caption{As in Figure~S15, but for the Superstorm Sandy ensembles.}
\label{fig:ageostrophic_sandy}
\end{figure}

\begin{figure}
\centering
\includegraphics[width=0.82\textwidth]{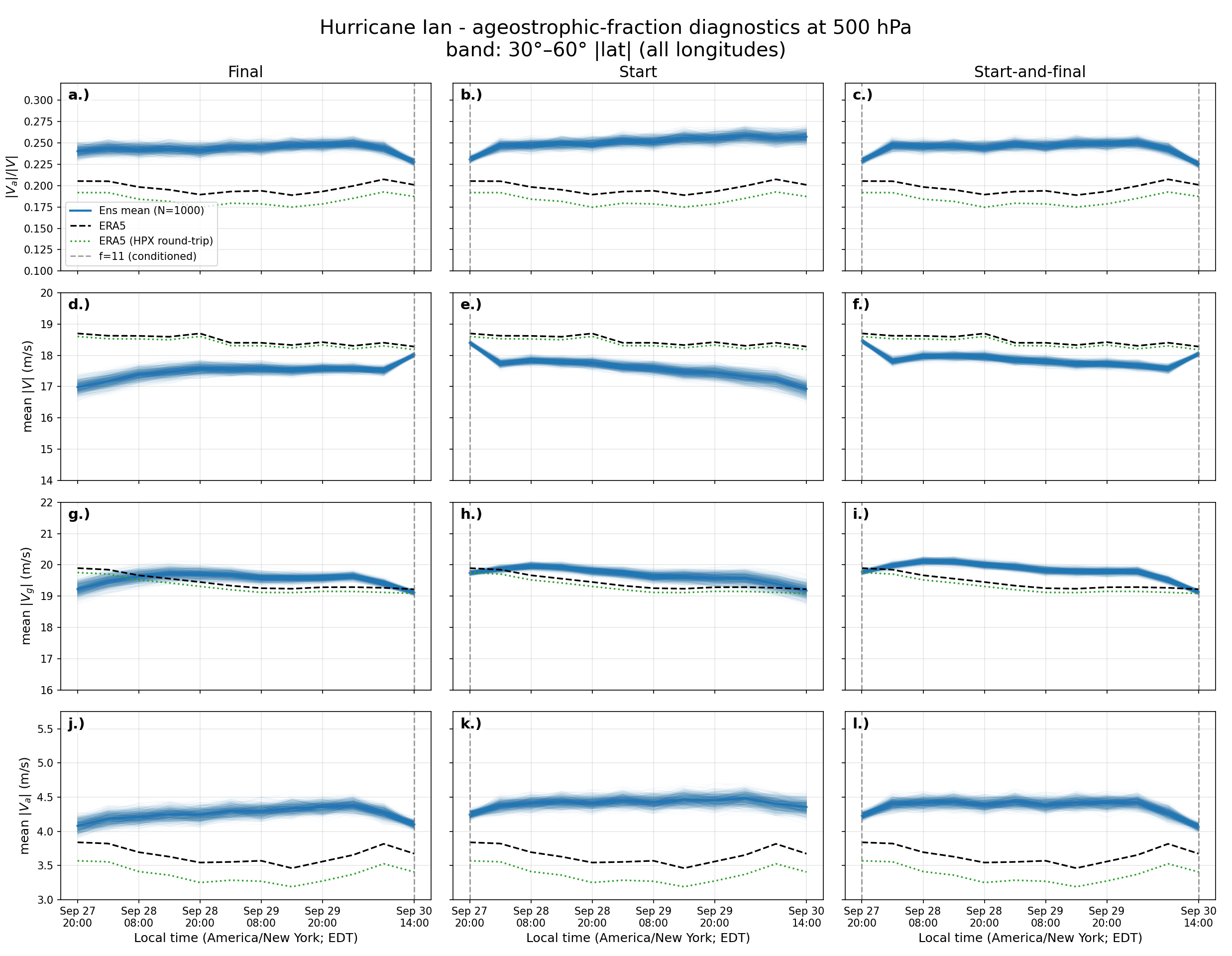}
\caption{As in Figure~S15, but for the Hurricane Ian ensembles. The $30^{\circ}$--$60^{\circ}$~$|\phi|$ band excludes Ian's Florida-landfall latitudes by construction to preserve consistent cross-member spatial sampling, as detailed in Text~S5.}
\label{fig:ageostrophic_ian}
\end{figure}

\end{document}